\documentclass{article} % For LaTeX2e
\usepackage{iclr2027_conference,times}

\usepackage{amsmath,amsfonts,bm}

\def\eqref#1{equation~\ref{#1}}
\def\1{\bm{1}}

\DeclareMathAlphabet{\mathsfit}{\encodingdefault}{\sfdefault}{m}{sl}
\SetMathAlphabet{\mathsfit}{bold}{\encodingdefault}{\sfdefault}{bx}{n}

\usepackage[T1]{fontenc}
\usepackage{hyperref}
\usepackage{url}
\usepackage{graphicx}
\usepackage{subcaption}
\usepackage{booktabs}   % \toprule \midrule \bottomrule
\usepackage{graphicx}   % \resizebox for the wide results table
\usepackage{amsmath}    % \pm, \rightarrow in text
\usepackage{cleveref}
\usepackage{enumitem}
\usepackage{booktabs}
\usepackage{graphicx}
\usepackage{subcaption}
\usepackage{tabularx}
\usepackage{textcomp}
\usepackage{listings}
\usepackage{placeins}
\usepackage{cleveref}   % 放在 hyperref 之后
\lstdefinestyle{prompt}{basicstyle=\ttfamily\scriptsize, breaklines=true, breakatwhitespace=false,
                        columns=fullflexible, keepspaces=true, frame=single, framesep=4pt}

\title{OpenSocialJax: Evaluating Social Adaptation in Sequential Social Dilemmas with Open-Ended Environmental Dynamics}

\author{
Zihao Guo$^{1}$ \enspace
Meng Fang$^{2}$ \enspace
Joel Z. Leibo$^{4}$ \enspace
Yali Du$^{1,3}$ \\
$^{1}$King's College London, 
$^{2}$University of Liverpool, 
$^{3}$The Alan Turing Institute,
$^{4}$Google DeepMind \\
\texttt{\{zihao.1.guo, yali.du\}@kcl.ac.uk} \\
\texttt{meng.fang@liverpool.ac.uk},
\texttt{jzl@deepmind.com}
}

\iclrfinalcopy % Uncomment for camera-ready version, but NOT for submission.
\begin{document}

\maketitle
\lhead{Preprint}

\maketitle

\begin{abstract}

Sequential social dilemma benchmarks capture diverse social interactions, but rarely test whether agents can infer hidden environmental rules through repeated interaction while adapting to other agents.
We introduce \textbf{OpenSocialJax}, which extends Clean Up and Commons Harvest across an open-ended distribution of latent environment configurations, systematically varying spatial layouts and hidden rules governing tool crafting, payoffs, and resource dynamics. 
Each episode samples one configuration from tens of millions of possibilities and holds it fixed across five trials. 
We evaluate eight large language models, including three closed source models and five open source models.
Our evaluation tests whether models can discover hidden rules and adapt to other agents in homogeneous and mixed teams. We find that rule discovery often improves performance but does not reliably sustain cooperation, as the cleaning burden becomes concentrated among fewer agents in OpenCleanup and every team eventually exhausts the shared resource in OpenHarvest. In our mixed arena, one cooperative agent was able to sustain the public good in OpenCleanup at a cost to its own return, but could not prevent resource depletion in OpenHarvest.
Together, these results highlight the difficulty of adapting jointly to unfamiliar environmental dynamics and the changing behavior of other agents. Code and demo are available at \url{https://github.com/cooperativex/OpenSocialJax}.

\end{abstract}

\section{Introduction}

Developing agents that can use experience to adapt across diverse and unfamiliar tasks is an important challenge. In multi-agent settings, adaptation requires agents to infer not only how the environment operates, but also the behaviors, capabilities, and objectives of other agents.
This is particularly challenging in sequential social dilemmas, where individual and collective interests may conflict. Consequently, it is crucial to develop benchmarks that systematically vary hidden environmental rules and dynamics. Such benchmarks enable controlled evaluation of how agents adapt to unfamiliar environmental mechanisms and the social dilemmas they induce.

Existing environments have made substantial progress in evaluating agents in sequential social dilemmas. Melting Pot \citep{meltingpot1,meltingpot2} provides a diverse suite of social environments with procedurally generated elements, while SocialJax \citep{guo2026socialjax} offers efficient JAX implementations of widely used social dilemma environments. However, these benchmarks are not designed to vary the hidden environmental dynamics underlying a social dilemma or evaluate how agents discover and adapt to them over repeated interactions.

Prior open-ended and meta learning environments have substantially expanded the diversity of settings available for training and evaluating adaptive agents. XLand \citep{xland_deepmind} constructs a vast, procedurally generated multi-agent task space, while XLand-MiniGrid \citep{xland_mini} provides a JAX-based implementation of 2D XLand. Craftax \citep{matthews2024craftax} offers a JAX-based benchmark for open-ended reinforcement learning. However, these environments do not explicitly examine how adaptation to changing environmental rules interacts with the social dilemmas those rules induce.

From the perspective of agent adaptation, large language model (LLM) agents offer a promising means of studying how agents reason about and respond to unfamiliar interactive environments.
Prior work has shown that LLMs can use textual descriptions of game states to reason, plan, and select actions in interactive games \citep{wu2023spring,wang2023voyager}.
These capabilities may allow them to infer latent environmental rules, reason about the behaviors and objectives of other agents, and revise their strategies as new evidence becomes available \citep{meta2022human}.
Nevertheless, their ability to adapt when environmental rules and social dependencies vary jointly remains unclear.

\begin{figure}[t]
    \centering
    \includegraphics[width=\linewidth]{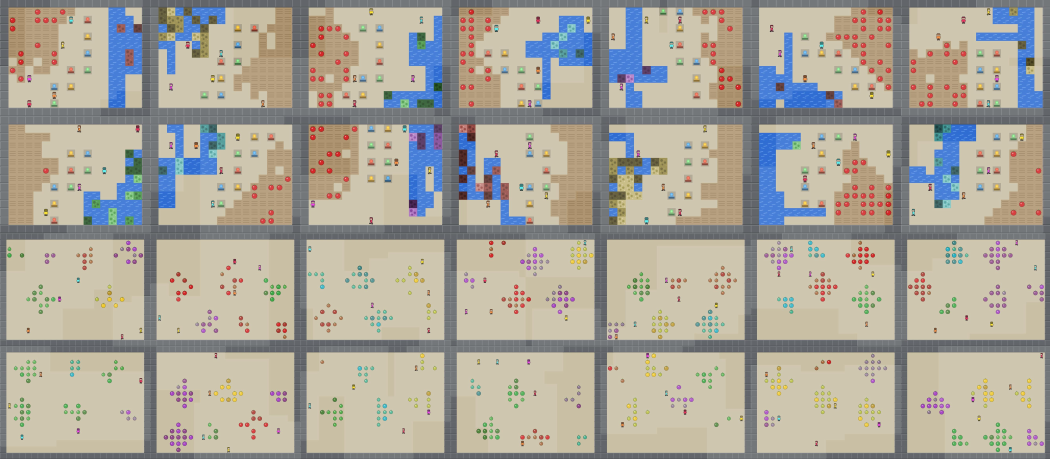}
    \vspace{-1.5em}
    \caption{Sampled task instances in OpenSocialJax. We show environment layouts for OpenCleanup (top two rows) and OpenHarvest (bottom two rows). Each episode instantiates a unique combination of spatial layouts (e.g., river topology, patch locations) and hidden rules governing tool crafting, rewards, and resource regeneration.}
    \label{fig:overview}
        \vspace{-1.0em}
\end{figure}

These gaps motivate benchmarks for open-ended experimentation that systematically vary hidden environmental rules and the social dependencies they induce. To address this need, we introduce OpenSocialJax, as shown in Figure~\ref{fig:overview}, which extends Clean Up and Commons Harvest \citep{meltingpot1,meltingpot2,guo2026socialjax} with newly designed procedural generators for environment layouts and hidden mechanisms. Each episode samples a latent configuration that remains fixed across multiple trials, allowing agents to accumulate evidence and revise their strategies through interaction. Resampling the configuration across episodes produces a broad and controllable distribution of unfamiliar conditions rather than a fixed collection of predefined scenarios. This design enables controlled investigation of how LLM-based agents explore unfamiliar settings, infer hidden mechanisms, and adapt to the evolving behaviors of other agents while navigating conflicts between individual and collective interests.

To evaluate how effectively LLMs reason and adapt in sequential social dilemmas, we consider eight recent models spanning diverse model families and capability levels. The closed source set includes GPT-6-Luna \citep{openai2026gpt6luna}, GPT-6-Sol \citep{openai2026gpt6sol}, and Claude Opus 5.5 \citep{anthropic2026opus55}. The open source set comprises Qwen3.8-27B \citep{qwen2026qwen38}, Qwen3.5-4B \citep{qwen2026qwen35}, DeepSeek-V4.1-Flash \citep{xu2026deepseek}, Gemma 4 \citep{team2026gemma}, and GLM-5.3-Flash \citep{zeng2026glm}. 
In summary, our contributions are threefold. First, we introduce OpenSocialJax, a benchmark of sequential social dilemmas whose map layouts and hidden rules are drawn from tens of millions of configurations. OpenSocialJax moves beyond dilemma benchmarks with fixed rules to test joint environmental and social adaptations. Second, we develop a protocol with repeated trials for evaluating LLM agents over long horizons, in which agents must adapt to both hidden environmental dynamics and the behavior of other agents. Agents maintain their own notes between trials, preserving their discoveries and enabling us to analyze how they respond to changing environmental dynamics in social dilemmas. Third, we evaluate eight recent open source and closed source frontier LLMs in both homogeneous populations and mixed arenas. We analyze the interplay between exploration guided by hypotheses, exploitation of the discovered rules, and free riding. Even the strongest model reasons explicitly about the dilemma in its own notes, yet still ends up depleting the shared resource or being exploited by others.

\section{Related Work}

\textbf{Open-ended and meta-learning environments.} 
Open-ended environments are important for studying agents adaptation and meta learning. 
There are several previous studies \citep{wang2016learningtoRL, duan2016rl^2, finn2017model_meta-learning, zintgraf2019varibad} which focus on tasks with simple distributions and policy adaptations. 
More benchmarks have since encompassed diverse robotic tasks \citep{yu2020meta-world}, latent causal structures \citep{wang2021alchemy}, survival environments \citep{matthews2024craftax}, and physics-based control tasks \citep{matthews2025kinetix}.
XLand \citep{xland_deepmind} constructs a large procedurally generated task space in which goals, objects, and environmental configurations. XLand-MiniGrid \citep{xland_mini} provides a lightweight JAX implementation of Xland.  
In multi-agent settings, XLand \citep{xland_deepmind} provides a procedurally generated task space spanning cooperative and competitive interactions. AdA \citep{team2023ada} studies rapid adaptation to novel tasks through interaction. MAESTRO \citep{samvelyan2023maestro} jointly constructs curricula over environmental parameters and co-player policies, while Multi-Agent Craftax \citep{matthews2024craftax} extends open-ended exploration to heterogeneous agents and complex cooperative tasks.

\textbf{Sequential dilemmas environments.}
Sequential social dilemma environments study temporally extended conflicts between individual and collective interests~\citep{leibo2017multi,perolat2017multi,hughes2018inequity}. Melting Pot provides a diverse suite of such environments~\citep{meltingpot1,meltingpot2}, while SocialJax offers efficient JAX implementations~\citep{guo2026socialjax}. 
JaxMARL \citep{rutherford2024jaxmarl} is a general-purpose MARL library that includes Coin Game and STORM, two grid-world environments that support the study of social dilemmas.
In contrast to these predefined environments and scenarios, OpenSocialJax varies hidden environmental mechanisms, enabling the study of adaptation across diverse latent social tasks.

\textbf{LLM agents in games.}
LLM-based agents have been evaluated in interactive games requiring reasoning, planning, exploration, and learning from experience~\citep{wu2023spring,wu2024smartplay,wang2023voyager,costarelli2024gamebench,paglieri2025balrog}. Other studies examine their cooperation, competition, negotiation, and deception in multi-agent games~\citep{meta2022human,chen2024llmarena,abdelnabi2024cooperation}. Recent work has also studied LLM-based agents in repeated games and social dilemmas, including matrix games, iterated Prisoner's Dilemma, public-goods games, intertemporal dilemmas, and sequential social dilemmas~\citep{akata2025playing,xu2024magic,willis2025will,piedrahita2025corrupted,nguyen2025navigating,gallego2026cooperation,tewolde2026coopeval}. Closest to our setting, GovSim \citep{piatti2024cooperate} shows that societies of LLM agents often deplete a shared resource, \citet{mosquera2025can} evaluate LLM-augmented agents in MeltingPot \citep{meltingpot2}, and Concordia \citep{vezhnevets2023generative} supports generative agent-based simulation of social interaction. However, how LLM agents infer latent game rules and adapt to varying environmental mechanisms and inter-agent dependencies remains underexplored.

\section{OpenSocialJax}

% We present OpenSocialJax, a benchmark suite for evaluating the environmental and social adaptation of LLM-based agents in sequential social dilemmas. 
% As one approach to evaluating open-ended adaptation, we use procedural generation to define a distribution of latent conditions for each environment by varying its layout and hidden environmental rules.
We present OpenSocialJax, a benchmark for evaluating environmental and social adaptation by LLM agents in sequential social dilemmas. To study open-ended adaptation, each environment procedurally varies its layout and hidden rules, creating a distribution of latent conditions.
% Each environment uses procedural generation to define a distribution of latent conditions by varying its layout and hidden environmental rules. 
Agents must infer unfamiliar game mechanisms through interaction and adjust their strategies in response to the observed behaviors of other agents.

\subsection{Problem Formulation}
\label{sec:difine}

\paragraph{Environment formulation.}
Each environment defines a distribution $p(z)$ over latent environment configurations. A configuration $z \sim p(z)$ induces a partially observable Markov game (POMG) \( \mathcal{G}_z = \left( \mathcal{N}, \mathcal{S}_z, \{\mathcal{A}_i\}_{i\in\mathcal{N}}, \{\mathcal{O}_i\}_{i\in\mathcal{N}}, \mathcal{P}_z, \{\Omega_{i,z}\}_{i\in\mathcal{N}}, \{\mathcal{R}_{i,z}\}_{i\in\mathcal{N}}, \rho_{0,z}, \gamma \right), \) where $\mathcal{N}$ is the set of agents, $\mathcal{S}_z$ is the state space, and $\mathcal{A}_i$ and $\mathcal{O}_i$ are the action and observation spaces of agent $i$. 
The transition function $\mathcal{P}_z(s' \mid s,\mathbf{a})$ determines the next state distribution given state $s$ and joint action $\mathbf{a}$, while $\Omega_{i,z}(o_i \mid s)$ defines the local observation received by agent $i$. The reward function $\mathcal{R}_{i,z}(s,\mathbf{a})$ specifies the reward of agent $i$, $\rho_{0,z}$ is the initial state distribution, and $\gamma$ is the discount factor. The latent configuration $z$ determines the environment layout and hidden mechanisms and may therefore affect the state space, transition dynamics, observations, rewards, and initial state distribution.

An episode samples one configuration $z \sim p(z)$ and consists of $K$ trials
\(
\mathcal{E}_z = \{\tau_z^1,\ldots,\tau_z^K\}.
\)
At the beginning of each trial, the environment state and agent positions are reset with the unchanged latent configuration $z$. The summarized interaction history available to each agent is retained across trials, allowing it to refine its understanding of the latent configuration.

\paragraph{Sequential social dilemmas under varying environments.}
For a fixed latent configuration $z$, let $\Pi_c^z$ and $\Pi_d^z$ denote the sets of cooperating and defecting policies. Consider a joint policy containing $l$ cooperating policies and $m$ defecting
policies, where $l+m=N$. We denote the corresponding average returns of cooperating and defecting policies by $R_c^z(l)$ and $R_d^z(l)$.

Following prior work \citep{leibo2017multi,hughes2018inequity}, the induced game
$\mathcal{G}_z$ constitutes a \emph{sequential social dilemma} when: (i) mutual cooperation is preferred to mutual defection, $R_c^z(N) > R_d^z(0)$; (ii) mutual cooperation is preferred to being exploited by defectors, $R_c^z(N) > R_c^z(0)$; and (iii) either the fear property or the greed property (or both) holds:
fear requires $R_d^z(i) > R_c^z(i)$ for sufficiently small $i$, whereas
greed requires $R_d^z(i) > R_c^z(i)$ for sufficiently large $i$.  The curves $R_c^z(l+1)$ and $R_d^z(l)$ form the Schelling diagram of $\mathcal{G}_z$, representing the expected returns from cooperating and defecting when $l$ co-players cooperate \citep{schelling1973hockey,perolat2017multi}. Across episodes, \(z\) varies the layout, hidden environmental mechanisms, and inter-agent dependencies while remaining unobserved by the agents. Agents must therefore infer the active configuration through interaction and adapt their strategies to both the environment and the observed behaviors of other agents.

\subsection{Environment Design}

Building on the Clean Up and Commons Harvest \citep{meltingpot1,meltingpot2, guo2026socialjax}, we develop two procedurally generated environments: OpenCleanup and OpenHarvest. 
Each environment combines generated layouts with hidden rule systems that can produce diverse environmental mechanisms and inter-agent dependencies. The active configuration is not revealed to the agents and must be inferred through interaction. Because agents adapt concurrently, their behaviors also change the social conditions faced by others. Agents must therefore adapt to both unfamiliar environmental mechanisms and the evolving behaviors of their agents.

% \begin{figure}[t]
%     \centering
%     \includegraphics[width=\linewidth]{figures/maps_structures/cleanup_discovery_structure_2026-09-13.pdf}
%     \caption{Overview of OpenSocialJax.}
%     \label{fig:overview}
% \end{figure}

% \begin{figure}[t]
%     \centering
%     \includegraphics[width=\linewidth]{figures/maps_structures/harvest_rule_figure_2026-09-19.pdf}
%     \caption{Overview of OpenSocialJax.}
%     \label{fig:overview}
% \end{figure}

\begin{figure}[t]
    \centering

    \begin{subfigure}{\linewidth}
        \centering
        \includegraphics[width=\linewidth]
        {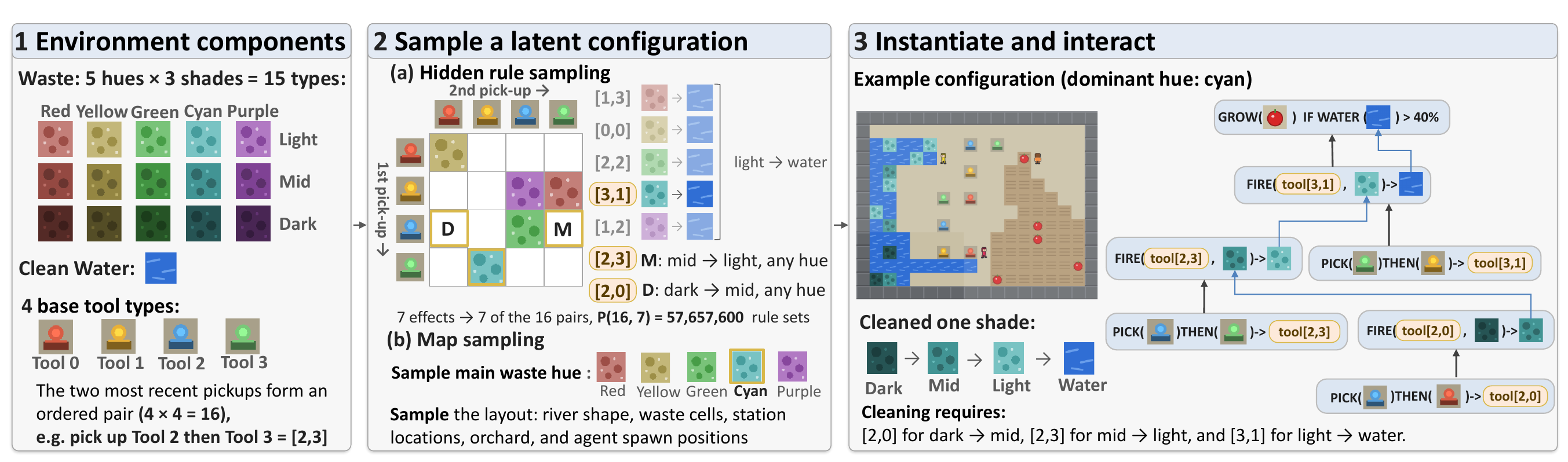}
        \caption{Overview of OpenCleanup.
(1) Five waste hues, three shades, and four base tools define the environment components and 16 possible ordered tool pairs.
(2) Seven cleaning effects are assigned to different pairs, while a dominant waste hue and map layout are sampled to form a latent configuration.
(3) Agents must infer the sampled tool rules and progressively clean waste from dark to mid, light, and water to enable apple growth.
The sampled configuration remains fixed across trials within an episode.}
        \label{fig:cleanup-structure}
    \end{subfigure}

    \vspace{0.5em}

    \begin{subfigure}{\linewidth}
        \centering
        \includegraphics[width=\linewidth]
        {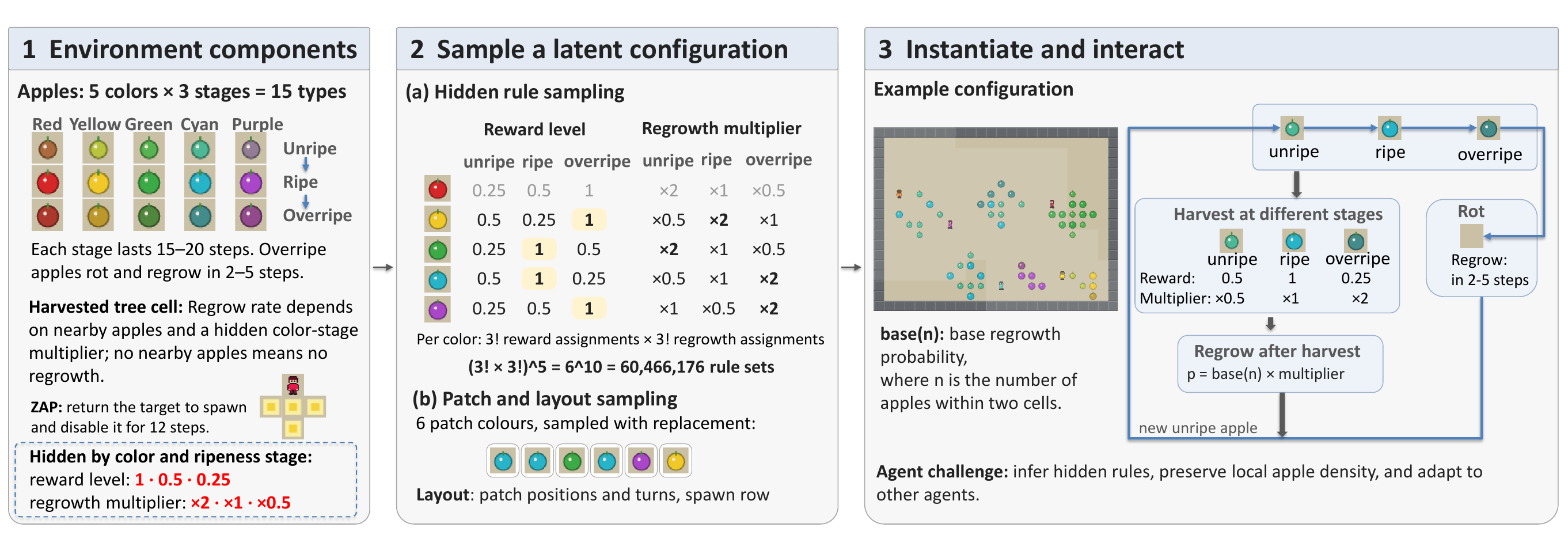}
        \caption{Overview of OpenHarvest.
(1) Apples vary across five colors and three ripeness stages.
(2) Hidden reward and regrowth rules, patch colors, and the map layout are sampled.
(3) In the instantiated environment, harvested cells regrow according to local apple density and a hidden multiplier, whereas rotted apples regrow independently.}
        \label{fig:harvest-rules}
    \end{subfigure}
\vspace{-1.5em}
    \caption{Overview framework of OpenSocialJax.}
    \label{fig:environment-overview}
    \vspace{-1.5em}
\end{figure}

\textbf{OpenCleanup.}
OpenCleanup retains the public goods dilemma of Clean Up. Agents receive rewards only by collecting apples, but apples grow only when the shared river is sufficiently clean. We extend the environment with a hidden tool system whose crafting rules and effects vary across episodes. Figure~\ref{fig:cleanup-structure} illustrates the environment components, the sampling of a latent configuration, and its instantiation during interaction.

Each waste cell has one of five hues and one of three shades, producing 15 visually distinct waste types. Dark waste must first be converted to mid waste, then to light waste, before it can be cleared into water. All three shades contribute to river pollution. Agents collect base tools from four types of tool stations. Their two most recent pickups form an ordered pair, with both order and repetition allowed, producing 16 possible pairs. Seven of these pairs are assigned distinct cleaning effects. Five are hue-specific cleaners, each clearing light waste of one particular hue; one converts mid waste of any hue to light waste; and one converts dark waste of any hue to mid waste. The remaining nine pairs have no effect. The seven effects are assigned to different tool pairs without replacement. The hidden tool system therefore contains $P(16,7) = \frac{16!}{9!} = 57{,}657{,}600$ possible rule sets. At the beginning of each episode, a rule set, a dominant waste hue, and a map layout are sampled and remain fixed across all trials. These conditions are hidden from the agents, who must explore the map, construct candidate tool pairs, fire them at waste, and infer their effects.

\textbf{OpenHarvest.}
OpenHarvest retains the common pool resource dilemma of Commons Harvest. Agents obtain rewards by consuming apples, but the probability of regrowth depends on how many apples remain nearby. This creates a tension between immediate individual rewards and sustaining the shared resource over time.  Figure~\ref{fig:harvest-rules} illustrates the environment components, the sampling of a latent configuration, and its instantiation during interaction.

We extend the original environment by introducing apple colors and visible ripeness stages, together with hidden reward and regrowth rules conditioned on both. OpenHarvest contains five possible apple colors and six orchard patches. At the beginning of each episode, the color of each patch is sampled independently with replacement, allowing the same color to appear in multiple patches. The spatial arrangement of the patches and the agent spawn locations are also randomly generated. Each apple progresses through three visible stages: unripe, ripe, and overripe, after which it rots and disappears. After an apple is consumed, its empty tree cell can regrow only while other apples remain nearby, with more neighboring apples producing faster regrowth. In contrast, a tree whose apple rots regrows after a fixed delay regardless of nearby apples. The sampled patch colors, layout, and hidden rules remain fixed across all trials within the episode. For each of the five possible colors, OpenHarvest independently assigns three reward levels and three regrowth levels to the three ripeness stages. The reward level determines the reward obtained by consuming an apple, while the regrowth level scales the base regrowth rate determined by nearby apples. Because the two assignments are sampled independently, the stage providing the highest immediate reward need not produce the fastest regrowth. Each assignment has \(3!=6\) possible orderings, giving
\(
(3!)^{2\times5}
=
6^{10}
=
60{,}466{,}176
\)
possible hidden rule sets.

\subsection{LLM Agent Interaction Protocol}
\label{sec:protocol}

We design the protocol so that agents must discover the game from their own experience and can improve over repeated play. Each agent can never see the hidden rule, the other agents' prompts or scores, and it cannot communicate with them. Instead of a growing conversation history, each agent relies on two kinds of memory. The harness keeps an objective record of what the agent has experienced, such as the effects of its own actions, while the agent keeps free-form notes in which it states its own conclusions and plans. Because the map and the hidden rule stay the same across the five trials of an episode, this memory lets agents accumulate evidence and revise their strategy from one trial to the next.

At every step, each agent is queried independently with a fixed system prompt, which explains the rules and the reply format, and a user message built by the harness. The user message contains the agent's egocentric $11\times11$ view, the actions available to it with their immediate effects, the rule evidence from its own experience, its recent steps, what it has seen other agents do, summaries of earlier trials, and its notes. The model replies with a JSON object that specifies one action and at most one edit to the notes; an invalid reply leaves the agent in place for that step. When a trial ends, the model rewrites its notes into a short list of lessons, and the environment is reset on the same map and rule. Only the rule evidence, the trial summaries and these lessons carry over to the next trial. Full prompts, the reply format and model settings are given in Appendices~\ref{app:harness} and~\ref{app:models}.

\section{Experiments}
\label{sec:experiments}

\subsection{Experimental setup}
\label{sec:exp-setup}

We evaluate three closed source models, GPT-6-Luna \citep{openai2026gpt6luna}, GPT-6-Sol \citep{openai2026gpt6sol} and Claude Opus 5.5 \citep{anthropic2026opus55}, and five open-source models, Qwen3.8-27B \citep{qwen2026qwen38}, Qwen3.5-4B \citep{qwen2026qwen35}, DeepSeek-V4.1-Flash \citep{xu2026deepseek}, Gemma 4 31B \citep{team2026gemma} and GLM-5.3-Flash \citep{zeng2026glm}, each with fixed inference settings (Appendix~\ref{app:models}). Our primary evaluation uses homogeneous teams of five instances of the same model. For each environment, every model is evaluated on 3 seeds; each seed samples one latent configuration and defines one episode of five trials of at most 400 steps, and an OpenHarvest trial ends early once no apple remains. We also run a mixed-model arena with Claude Opus 5.5, GPT-6-Sol, GLM-5.3-Flash, DeepSeek-V4.1-Flash and Qwen3.8-27B, using the same seeds, prompts and inference settings. Appendix~\ref{app:models} summarizes the settings and configurations of experiments.

\subsection{Main Results}

\begin{figure*}[t]
    \centering
    \begin{subfigure}[b]{0.245\textwidth}
        \centering
        \includegraphics[width=\linewidth]{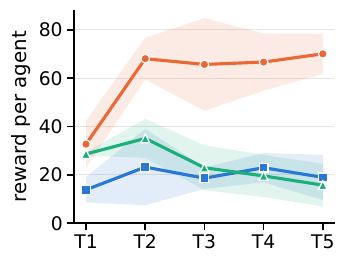}
        \caption{OpenCleanup, closed}
        \label{fig:results_cleanup_closed}
    \end{subfigure}\hfill
    \begin{subfigure}[b]{0.245\textwidth}
        \centering
        \includegraphics[width=\linewidth]{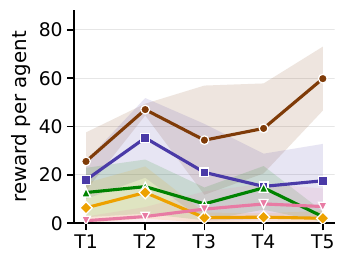}
        \caption{OpenCleanup, open}
        \label{fig:results_cleanup_open}
    \end{subfigure}\hfill
    \begin{subfigure}[b]{0.245\textwidth}
        \centering
        \includegraphics[width=\linewidth]{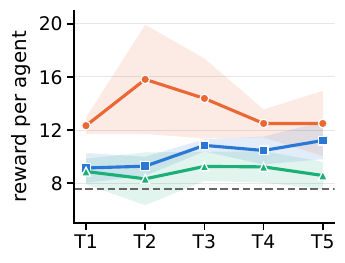}
        \caption{OpenHarvest, closed}
        \label{fig:results_harvest_closed}
    \end{subfigure}\hfill
    \begin{subfigure}[b]{0.245\textwidth}
        \centering
        \includegraphics[width=\linewidth]{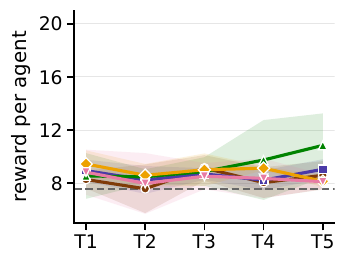}
        \caption{OpenHarvest, open}
        \label{fig:results_harvest_open}
    \end{subfigure}

    \vspace{2pt}
    \includegraphics[width=1\textwidth]{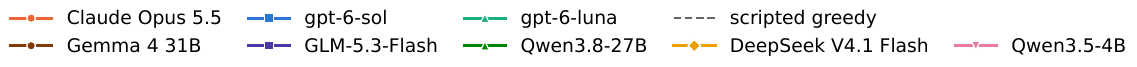}

    \caption{Mean per-agent reward per trial for closed source (closed) and open source (open) models. Shaded bands: $\pm1$ s.d.\ across seeds. Dashed line: scripted greedy policy.}
    \label{fig:main_results}
\end{figure*}

We first examine how effectively homogeneous populations perform and adapt across repeated trials. Figure~\ref{fig:main_results} reports the average agent return over the five trials in OpenCleanup and OpenHarvest. For each configuration, returns are first averaged across the five agents, and each curve then shows the mean across different seeds, with shaded regions indicating standard deviation. Because the latent configuration remains fixed while the environment resets between trials, changes in return indicate whether agents can use experience from earlier trials to improve their subsequent behavior. This comparison captures both initial performance under unfamiliar conditions and adaptation through repeated interaction. 
Claude Opus 5.5 is the strongest model in both environments.
In OpenCleanup, as illustrated in \Cref{fig:results_cleanup_closed,fig:results_cleanup_open}, its return jumps after the first trial and stays at that level for the remaining trials.
Almost every model improves from the first to the second trial, but for several of them the gain does not last. Their return peaks in the second trial and then drifts back down. 
In OpenHarvest, as illustrated in \Cref{fig:results_harvest_closed,fig:results_harvest_open}, every model performs better than the scripted greedy policy (hand-crafted greedy policy that can eat apples as soon as possible), but only Claude Opus 5.5 and gpt-6-sol clearly separate from it. 

% preamble 里需要：
% \usepackage{graphicx}
% \usepackage{subcaption}

\begin{figure*}[t]
    \centering
    \begin{subfigure}[b]{0.245\textwidth}
        \centering
        \includegraphics[width=\linewidth]{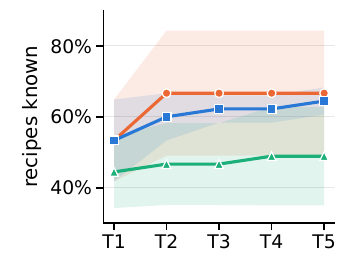}
        \caption{OpenCleanup, closed}
        \label{fig:knowledge_cleanup_closed}
    \end{subfigure}\hfill
    \begin{subfigure}[b]{0.245\textwidth}
        \centering
        \includegraphics[width=\linewidth]{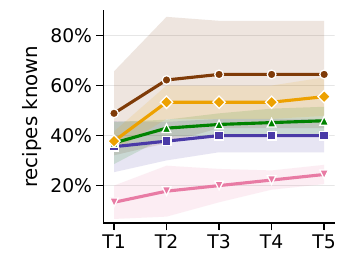}
        \caption{OpenCleanup, open}
        \label{fig:knowledge_cleanup_open}
    \end{subfigure}\hfill
    \begin{subfigure}[b]{0.245\textwidth}
        \centering
        \includegraphics[width=\linewidth]{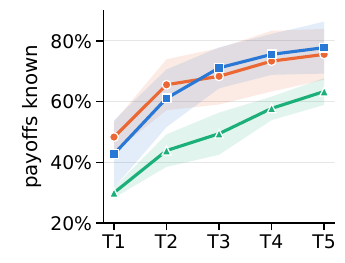}
        \caption{OpenHarvest, closed}
        \label{fig:knowledge_harvest_closed}
    \end{subfigure}\hfill
    \begin{subfigure}[b]{0.245\textwidth}
        \centering
        \includegraphics[width=\linewidth]{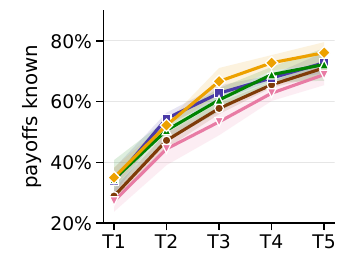}
        \caption{OpenHarvest, open}
        \label{fig:knowledge_harvest_open}
    \end{subfigure}

    \vspace{2pt}
    \includegraphics[width=1\textwidth]{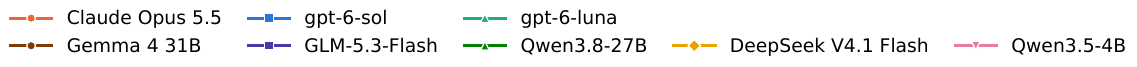}

    \caption{Hidden rules an agent knows by the end of each trial, averaged over agents.}
    \label{fig:knowledge}
    \vspace{-1.0em}

\end{figure*}

% \paragraph{What agents learn.}
\Cref{fig:knowledge} tracks how much of the hidden rules each agent has discovered.
In OpenCleanup, agents learn nearly all the recipes they will ever know in the first two trials; after that, the curves are flat.
No model's agents learn every recipe, because an agent only learns a recipe from its own shots. Moreover, an individual agent does not need to discover all hidden rules to perform well, since agents can collectively solve the task by specializing in different subsets of the rules. For example, agents can specialize in different tools and divide the work of cleaning the river. This behavior is also evident in the agents’ trajectories: Claude Opus 5.5 observes which tools other agents are using and then selects its own tool to complement their actions. In OpenHarvest, by contrast, knowledge of the color–ripeness payoffs continues to grow across all five trials, and most models end up knowing a similar share. Interestingly, discovering the hidden rules does not necessarily mean that agents use them effectively. \Cref{tab:app_use} (Appendix~\ref{app:use}) reports the details of whether agents can use these hidden rules to get more reward.

\begin{figure*}[t]
    \centering
    \begin{subfigure}[b]{0.32\textwidth}
        \centering
        \includegraphics[width=\linewidth]{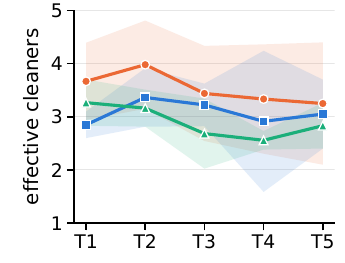}
        \caption{OpenCleanup: cleaners, closed}
        \label{fig:dilemma_cleaners_closed}
    \end{subfigure}\hfill%
    \begin{subfigure}[b]{0.32\textwidth}
        \centering
        \includegraphics[width=\linewidth]{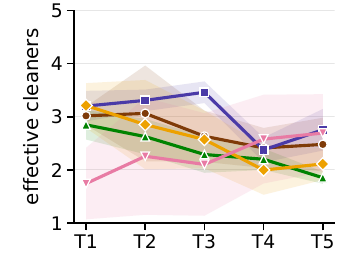}
        \caption{OpenCleanup: cleaners, open}
        \label{fig:dilemma_cleaners_open}
    \end{subfigure}\hfill%
    \begin{subfigure}[b]{0.32\textwidth}
        \centering
        \includegraphics[width=\linewidth]{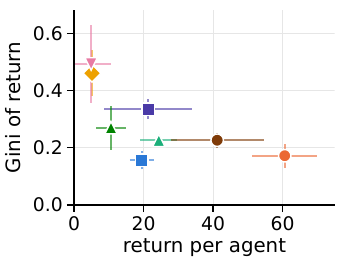}
        \caption{OpenCleanup: return vs.\ Gini}
        \label{fig:dilemma_cleanup_gini}
    \end{subfigure}

    \vspace{4pt}
    \begin{subfigure}[b]{0.32\textwidth}
        \centering
        \includegraphics[width=\linewidth]{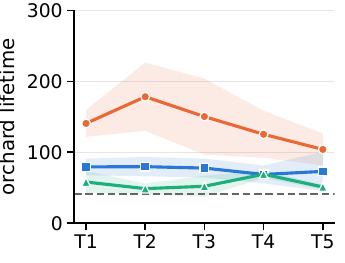}
        \caption{OpenHarvest: orchard, closed}
        \label{fig:dilemma_orchard_closed}
    \end{subfigure}\hfill%
    \begin{subfigure}[b]{0.32\textwidth}
        \centering
        \includegraphics[width=\linewidth]{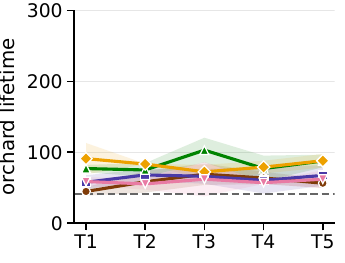}
        \caption{OpenHarvest: orchard, open}
        \label{fig:dilemma_orchard_open}
    \end{subfigure}\hfill%
    \begin{subfigure}[b]{0.32\textwidth}
        \centering
        \includegraphics[width=\linewidth]{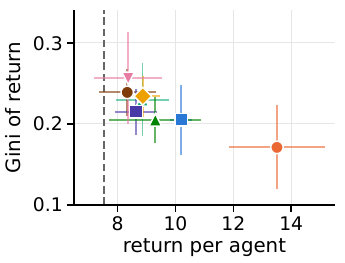}
        \caption{OpenHarvest: return vs.\ Gini}
        \label{fig:dilemma_harvest_gini}
    \end{subfigure}

    \vspace{2pt}
    \includegraphics[width=1\textwidth]{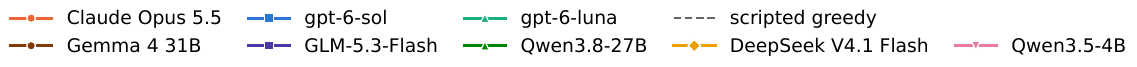}

    \caption{Social dilemma analysis.
(a, b) Effective number of cleaners in each OpenCleanup trial, defined as $1/\sum_i s_i^2$, where $s_i$ is agent $i$'s share of the team's total cleaning contribution, measured by the number of waste tiles cleared.
    A value of 5 indicates that all five agents contribute equally, whereas a value of 1 indicates that a single agent performs all the cleaning.
   (d, e) Orchard lifetime in each OpenHarvest trial, measured as the timestep at which the final apple is eaten.
    (c, f) Mean per-agent return against the Gini coefficient of returns within a trial, averaged over trials; bars show $\pm1$ s.d.\ across seeds, and the dashed line in (f) is the scripted greedy policy.}
    \label{fig:dilemma}
        \vspace{-1.5em}
\end{figure*}

\subsection{Social Dilemma Analysis}
\label{sec:dilemma}

In both environments, agents observe and optimize only their own rewards, while their actions also affect the opportunities available to others. This mismatch between individual incentives and collective outcomes gives rise to two social dilemmas, which we verify using the Schelling diagrams in Appendix~\ref{app:schelling}. OpenCleanup presents a public-goods dilemma, where cleaning is individually costly but enables apple growth that benefits all agents. OpenHarvest presents a common resource dilemma, where harvesting yields immediate individual rewards but excessive consumption reduces future regrowth for everyone. \Cref{fig:dilemma} examines how agents navigate these tensions and how evenly rewards are distributed among them.

\textbf{Contribution to the public good.}
In OpenCleanup (\Cref{fig:dilemma_cleaners_closed,fig:dilemma_cleaners_open}), we measure how widely the cleaning is shared within a team using the effective number of cleaners.
For nearly every model, this number decreases across trials, indicating that cleaning becomes concentrated among fewer agents while the others increasingly focus on harvesting.
Total cleaning separates specialization from free-riding, as shown in Appendix~\ref{app:total-cleaning}. Claude Opus 5.5 and Gemma 4 31B keep cleaning and returns high, whereas GPT-6-Sol, GPT-6-Luna, Qwen3.8-27B and DeepSeek-V4.1-Flash clean less in later trials and earn less.
The agents’ notes help explain how this shift toward free-riding develops. When agents observe that others are cleaning, they may conclude that the river can remain productive without their own contribution.
After a successful trial, some agents explicitly decide to rely on others to clean. As one agent notes, \emph{“I can rely on other agents’ cleanup while focusing on harvesting.”} Because agents see only their own apple rewards, they may overlook how much the team’s cleaning contributed to those rewards.
As a result, they may attribute the high reward primarily to harvesting and continue to free-ride in later trials. Claude Opus 5.5 maintains the broadest participation in cleaning, although it follows the same downward trend. Qwen3.5-4B is the only exception. Its increase appears to reflect delayed mastery of the tool recipes rather than growing cooperation.

\textbf{Sustaining the common resource.}
In OpenHarvest (\Cref{fig:dilemma_orchard_closed,fig:dilemma_orchard_open}), we measure how long the orchard survives: the step at which the last apple is eaten, after which nothing can regrow.
No model keeps the orchard alive for a full trial.
The open-source models and the GPT models strip it early in every trial, and their orchard lifetime does not improve with experience, which matches their returns staying close to the greedy baseline.
Claude Opus 5.5 keeps the orchard productive for much longer by correctly inferring that it should wait for apples to regrow. 
Its orchard lifetime peaks in the second trial and then shortens (\Cref{fig:dilemma_orchard_closed}).
Its agents' notes show why. From the third trial on, almost all of them conclude that holding back does not pay because the others strip the apple patches anyway, for example \emph{``Saving apples for regrowth rarely pays; eat the good ones before rivals do''}.

\textbf{Efficiency and equality.}
The third column (\Cref{fig:dilemma_cleanup_gini,fig:dilemma_harvest_gini}) plots each model's mean return against the Gini coefficient of returns within a trial.
In both environments, Claude Opus 5.5 combines the highest return with one of the most even distributions, so efficiency does not come at the cost of fairness.
In OpenCleanup, gpt-6-sol distributes returns just as evenly but at a much lower level. The models with the lowest returns show the most unequal distributions, although the Gini coefficient is less reliable when few apples are eaten.
In OpenHarvest, all models except Claude Opus 5.5 cluster near the greedy baseline with similar inequality.

Current frontier models show promise in navigating social dilemmas, but
none of the models fully resolve either dilemma.
In OpenCleanup, agents initially contribute actively to river cleaning, but across successive trials they increasingly learn to free-ride, leaving progressively fewer agents to bear the cleaning burden. 
In OpenHarvest, the shared resource is exhausted within each trial.
Stronger models delay these failures rather than avoid them.

\label{sec:arena}

\begin{figure*}[t]
    \centering
    \begin{subfigure}[b]{0.245\textwidth}
        \centering
        \includegraphics[width=\linewidth]{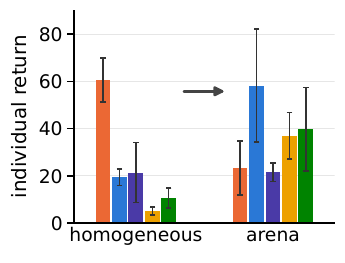}
        \caption{OpenCleanup: return}
        \label{fig:arena_cleanup_return}
    \end{subfigure}\hfill%
    \begin{subfigure}[b]{0.245\textwidth}
        \centering
        \includegraphics[width=\linewidth]{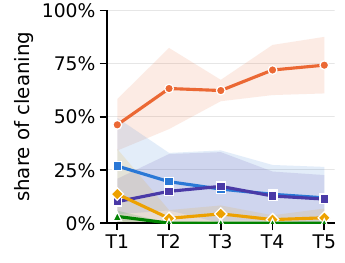}
        \caption{OpenCleanup: cleaning}
        \label{fig:arena_cleanup_share}
    \end{subfigure}\hfill%
    \begin{subfigure}[b]{0.245\textwidth}
        \centering
        \includegraphics[width=\linewidth]{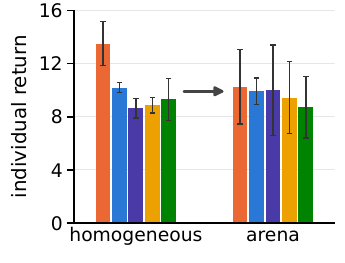}
        \caption{OpenHarvest: return}
        \label{fig:arena_harvest_return}
    \end{subfigure}\hfill%
    \begin{subfigure}[b]{0.245\textwidth}
        \centering
        \includegraphics[width=\linewidth]{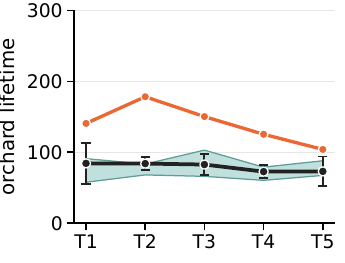}
        \caption{OpenHarvest: orchard}
        \label{fig:arena_harvest_orchard}
    \end{subfigure}

    \vspace{2pt}
    \includegraphics[width=1\textwidth]{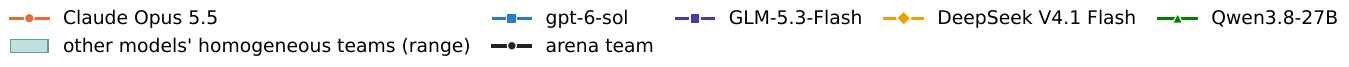}

    \caption{OpenSocialJax arena: each model controls one of the five agents, with seats rotated across seeds.
(a, c) Individual return of each model in a homogeneous team and in the arena.
    (b) Each model's share of its team's cleaning in the arena.
(d) Orchard lifetime of the arena team, compared with homogeneous teams: Claude Opus 5.5 and, shaded in teal, the range covered by the other four arena models' homogeneous teams.}
    \label{fig:arena}
            \vspace{-1.5em}

\end{figure*}

\subsection{OpenSocialjax Arena}

We next evaluate the models in a mixed-model arena. Five different models play in the same game, with each controlling one of the five agents, as shown in \Cref{fig:arena}.
We compare each model’s behavior and return in the arena with those in a homogeneous team to examine the effect of heterogeneous co-players.

\textbf{Public goods: the cooperator pays.}
In OpenCleanup, Claude Opus 5.5 keeps cleaning the river in the arena, and its share of the team's cleaning grows over the trials (\Cref{fig:arena_cleanup_share}).
Meanwhile, DeepSeek-V4.1-Flash and Qwen3.8-27B almost stop cleaning.
Claude's agent notices this and keeps cleaning anyway, writing for example \emph{``D never cleans. My best role is clearing G0.''}
Claude bears most of the cost. Its return falls far below that achieved in the homogeneous Claude Opus 5.5 team, while the models that contribute least to cleaning gain the most (\Cref{fig:arena_cleanup_return}).
The public good is sustained, and the mixed-model team collects more apples than any homogeneous team except the Claude Opus 5.5 team.
A single cooperative agent can therefore sustain the public good, but doing so reduces its own interest.

\textbf{Common resources: the cooperator gives up.}
OpenHarvest shows the opposite outcome.
Claude Opus 5.5’s advantage disappears in the mixed-model arena. Its return falls to the level of the other models, while some models that perform worse in homogeneous teams achieve slightly higher returns in the arena (\Cref{fig:arena_harvest_return}).
The mixed team depletes the orchard as quickly as the homogeneous teams composed of the other models, and far earlier than the homogeneous Claude Opus 5.5 team (\Cref{fig:arena_harvest_orchard}).
Claude's notes show why. It soon learns that restraint no longer pays when others strip the patches: \emph{``Apples held back mostly go to others''} and \emph{``Leaving apples there to regrow is futile. Eat fast.''}
It still picks apples at their most valuable stage, but it no longer leaves them to regrow.
One restrained agent cannot protect a shared resource from four agents that deplete it.

% \paragraph{Two dilemmas, two responses.}
% The same model responds differently to the two dilemmas.
% In OpenCleanup, nobody eats if nobody cleans, so continuing to clean still pays for Claude, even while others free-ride.
% In OpenHarvest, restraint only leaves more apples for others, so Claude stops holding back once they do not.
% In both cases, the collective outcome of a mixed team depends on its least cooperative members.
% A strong model can make up for them in a public-goods game, at a cost to itself, but not in a common-pool resource game.
% Because each model controls a single agent and we run three seeds, these results are qualitative; we report per-seed values in \Cref{app:arena}.

\section{Conclusion}
In this work, we introduced \textbf{OpenSocialJax}, an environment suite for evaluating how multi-agent systems adapt to unfamiliar environmental dynamics in sequential social dilemmas. \textbf{OpenSocialJax} extends Clean Up and Commons Harvest with procedurally generated layouts and hidden rules governing tool effects, rewards, and resource dynamics, producing tens of millions of possible configurations. We evaluated eight proprietary and open source language models in homogeneous teams and a mixed model arena. The models discovered many hidden rules and often improved after the first trial, but rule discovery did not reliably lead to sustained cooperation. Across trials, cleaning became concentrated among fewer agents in OpenCleanup, while every team eventually depleted the shared resource in OpenHarvest. These results show the difficulty of adapting simultaneously to unfamiliar environmental dynamics and the changing behavior of other agents.

\textbf{Limitations and Future Work.}
OpenSocialJax is implemented in JAX for fast parallel simulation, but training learning-based agents that adapt to the latent configuration remains an open challenge. Future work could extend the multi-trial protocol with explicit communication and more sophisticated theory of mind mechanisms to study how agents resolve social dilemmas under uncertainty. We hope \textbf{OpenSocialJax} provides a foundation for developing agents that discover environmental rules, anticipate collective behavior, and sustain cooperation in unfamiliar social environments.

\section*{AI Use Statement}
LLMs are the subject of this study: the eight models listed in Section~\ref{sec:experiments} are evaluated as agents, and all agent behavior reported in this paper was generated by them. Separately, we used GPT and Claude to assist with manuscript drafting, editing and finding related work. We also used Claude Code to assist with code editing, refactoring, and debugging. We have not used generative AI tools to generate or modify experimental results or agent transcripts. All AI-assisted content was reviewed, checked against our data and code, and revised by the authors.

\section*{Reproducibility Statement}
\textbf{Code and Data Availability.} The LLM agents prompts and harness are available at the repository linked in the abstract. All experimental details and configurations are provided in the repository, and the prompts used in our experiments are included in the Appendix~\ref{app:harness}.

\bibliography{iclr2027_conference}
\bibliographystyle{iclr2027_conference}

% \newpage

\appendix
% \Crefalias{section}{appendix}
% \Crefname{appendix}{Appendix}{Appendices}
% \Crefname{appendix}{Appendix}{Appendices}
% ===================================================================================================
% Appendix for the LLM-agent social-dilemma experiments.
% Upload the folder drafts/appendix/ to the Overleaf project as appendix/ and use \input{appendix/appendix}
% after \appendix. Tables and figures are generated by drafts/analysis_scripts/gen_appendix.py.
% Preamble needs: booktabs, graphicx, listings, tabularx, textcomp (\textperthousand), cleveref, placeins (\FloatBarrier).
%   \lstdefinestyle{prompt}{basicstyle=\ttfamily\scriptsize, breaklines=true, breakatwhitespace=false,
%                           columns=fullflexible, keepspaces=true, frame=single, framesep=4pt}
% TODO markers: statements taken from an older version of the code (github branch feat/xland-style-task,
% 2026-09-22) that could not be checked against the logged runs. Verify them against the run code.
% ===================================================================================================
\providecommand{\FloatBarrier}{}   % from placeins; keeps floats inside their section

\section{LLM Agent}
\label{app:harness}

Every agent is an independent call to the same model. At each step, each agent receives the system prompt of its environment (Listings~\ref{lst:system_cleanup} and~\ref{lst:system_harvest}) and one user message that describes the current state from its own point of view (Listings~\ref{lst:step_cleanup} and~\ref{lst:step_harvest}). The model replies with one JSON object that contains its action. Agents do not talk to each other, and a model never sees another agent's messages, notes or scores. Agents can only see others' actions when they in the observation space.
% TODO verify: no chat history is kept between steps (each call is [system, user]).

\paragraph{Observation.}
The user message starts with a progress line (trial, step, and the agent's own score so far in the trial), followed by these blocks:
\begin{itemize}
    \item \textbf{RULES YOU KNOW}: what the agent has established from its own experience, maintained by the harness. In OpenCleanup, this is every pick-up order the agent has fired with and the effect of those shots on each waste color. In OpenHarvest, it lists the points paid by each color and ripeness stage the agent has eaten, the regrowth times of cells it saw emptied, and the ripening and rotting times it observed. This block persists for the whole episode.
    \item \textbf{YOUR NOTES}: free-form notes written by the agent itself (see below).
    \item \textbf{EARLIER TRIALS}: one summary per finished trial, written by the harness: what the agent did and what it saw of the others.
    \item \textbf{OTHER AGENTS}: where each other agent was last seen and what it was doing, limited to the agent's own view.
    \item \textbf{YOUR PLAN}: the agent's current and previous goals.
    \item \textbf{LAST STEP} and \textbf{RECENT STEPS}: the outcome of the last action, events in view, and the last 15 steps.
    \item \textbf{NOW}: position, facing and hand, and an $11\times11$ egocentric view drawn north-up with absolute coordinates. OpenHarvest adds the patches seen so far and the emptied cells the agent is waiting on.
    \item \textbf{ACTIONS NOW}: the available actions and what each would do (e.g.\ which cell a move lands on, or which waste a shot would hit).
\end{itemize}

\paragraph{Memory across steps and trials.}
Agents keep three kinds of memory. The harness keeps RULES YOU KNOW and EARLIER TRIALS up to date. The agent itself writes its notes, applying at most one edit per step (add an entry, replace or delete an entry by its number, or none). When a trial ends, the model is asked to rewrite its notes into a short list of \emph{lessons} and \emph{plans for the next trial}. Only this list is carried into the next trial; the step-by-step entries are dropped. If the rewrite fails, the agent starts the next trial without notes. The agents' own lessons are the source of the quotations in Appendix~\ref{app:notes}.

\paragraph{Frozen agents and early stopping in OpenHarvest.}
An agent hit by a zap is sent back to a spawn point and cannot act for 12 steps. During these steps, the model is not called and no prompt is recorded. When no apple is left on the map, no apple can appear again, since an empty cell only regrows next to hanging apples. The harness then skips the rest of the trial; this does not change any return.
% TODO verify the exact early-stop condition in the run code (older code: no apple left and no rotted cell waiting to regrow).

\begin{lstlisting}[style=prompt, caption={System prompt, OpenCleanup (as shown to agent A).}, label={lst:system_cleanup}]
You are agent A, one of 5 agents (A, B, C, D, E) in a grid world called Clean Up.

1. YOUR OBJECTIVE
1.1 Score = apples eaten. Only the apples YOU eat count for you. Cleaning the river helps everyone; apples go to whoever gets there first.
1.2 Apples grow only in the orchard, and only while the river is clean enough. Where the river, the orchard and the stations are changes from episode to episode and stays the same for all trials of an episode: find them by looking around.
1.3 A trial lasts 400 steps. Then the map is re-laid in the same layout (stations stay where they were): the river is fully polluted again, your hand is empty, apples are gone. The recipes (see 2.6) and what each working tool does stay the same for all 5 trials of the episode.
1.4 You are told only what YOU cleared and ate. You never see other agents' scores, hands or windows.

2. THE WORLD
2.1 Waste cells have a HUE (red, yellow, green, cyan, purple) and a SHADE (light, mid, dark). Colour code = hue*3 + shade, e.g. 6 = green-light, 7 = green-mid, 8 = green-dark.
2.2 Firing your beam at a waste cell does exactly one of: CLEARS it to water, makes it ONE SHADE LIGHTER (dark -> mid, mid -> light), or NOTHING.
2.3 A WORKING tool is made by picking up two base tools in order; a base tool on its own does nothing to waste. Which pick-up order makes which working tool, and which waste that tool acts on, you find out by trying it -- RULES YOU KNOW then keeps it for you.
2.4 There are 4 kinds of BASE tool stations (tools 0-3), in a chain between the river and the orchard, several of each.
2.5 PICK UP = action 8 while standing on a station. Walking onto or across a station does nothing to your hand.
2.6 You have one hand and a WINDOW = your last two pick-ups, oldest first. Every pick-up pushes the oldest one out: window [2, 0], pick up 3 -> window [0, 3]. Each working tool has a hidden RECIPE, an ordered pair such as [2, 0] = pick up tool 2, then tool 0. The moment your window equals a recipe, that working tool is in your hand; the next pick-up slides the window and you hold a plain base tool again. No two working tools have the same recipe.
2.7 The observation names what is in your hand only by the pick-ups that made it, never by what it is or does -- it does not even tell you whether those two pick-ups were a recipe at all. Firing is what tells you.
2.8 Firing: the beam covers the cell you FACE, the cell beyond it, and the two cells diagonally beside the faced cell.
2.9 Moves are absolute compass moves and do not change your facing; only turning does. A move into a wall or another agent is blocked. Walking onto an apple eats it (+1). Waste slowly reappears.

3. HOW TO READ YOUR TURN
3.1 Positions are (row, col): row 0 is the NORTH edge and grows southward; col 0 is the WEST edge and grows eastward. North = row-1, south = row+1, east = col+1, west = col-1.
3.2 You see an 11x11 window that turns with you: 9 cells ahead of you, 1 cell behind and 5 to each side, so turning changes what you see. The map shows exactly those cells, always drawn north-up with columns labelled on top and rows on the left (ME sits one cell in from the side behind you); cells beyond the edge of the world are ##. ME = you, @B = agent B (every other agent is shown by its letter), T<n> = station of tool n, AP = apple, ,, = orchard soil (apples grow there), ~~ = water, ## = wall, .. = ground, waste = hue letter + shade digit (G2 = green-dark).
3.3 RULES YOU KNOW: every pick-up order you have fired with and what those shots turned out to do, worked out and kept for you -- you never have to write a shot down yourself. YOUR NOTES: yours, and the only thing here that is -- what you have worked out about how to PLAY, and what you mean to do next. Nothing in it is checked or filled in for you. EARLIER TRIALS: one line per finished trial with what you did and what you saw of the others; use it to do better this trial. OTHER AGENTS: only what you have seen in your own view: where each agent was, whether it was firing, and what its shots did to the waste cells you could see. YOUR PLAN: your current goal and the goals you set before, kept for you. LAST STEP: what your last action did, and events that happened in your view. RECENT: your last few steps. ACTIONS NOW: what you can do right now; fire shows what the beam would hit.
3.4 You know only what you see: other agents and the river exist beyond your view, but you get no information about them.

4. HOW TO ANSWER
4.1 Reply with ONE JSON object and nothing else:
{"why_this_move": "<up to 500 characters>",
 "mode": "craft" | "clean" | "harvest" | "explore",
 "goal": {"type": "move_to" | "fire" | "pick_up" | "wait", "cell": [row, col], "why": "<up to 80 characters>"},
 "note": {"op": "none" | "add" | "replace" | "delete", "id": <entry number to replace or delete, else 0>, "text": "<the entry to add, or the new wording for [id]; \"\" otherwise>"},
 "action": <one integer from ACTIONS NOW>}
4.2 mode = what you are working on: craft (getting a working tool), clean (firing at waste), harvest (going for apples), explore (testing an unknown pair).
4.3 goal = the sub-goal your action serves. Keep the SAME goal from step to step until it is done or an event makes it pointless; if you change it, say why in "why". For fire / pick_up / wait, cell = where you do it.
4.4 Example, keeping the goal:
{"why_this_move": "Both shots I have fired since my window became [2, 0] turned a G2 cell into G1, which nothing I held before did. Another G2 is at (6,3), so this hand is worth keeping while I work on it. I am at (6,8).", "mode": "clean", "goal": {"type": "move_to", "cell": [6, 5], "why": "get within reach of the G2 cells"}, "note": {"op": "none", "id": 0, "text": ""}, "action": 3}
4.5 Example, changing the goal after an event:
{"why_this_move": "That shot changed nothing at all, so window [3, 1] leaves G0 alone - the third window I have fired at G0 with no effect. I am standing on T2 and have never held [2, 0]. Picking up 2 here starts it.", "mode": "explore", "goal": {"type": "pick_up", "cell": [6, 4], "why": "start an untried window [2, 0]"}, "note": {"op": "add", "id": 0, "text": "every order I test costs a walk to the stations and back, so fire at every shade within reach before going back"}, "action": 8}

5. YOUR NOTES
5.1 What your shots did is kept for you under RULES YOU KNOW, for the whole episode. You never have to write a shot down, and you should not: re-recording one only clutters your notes.
5.2 What is NOT kept for you is how to play. "note" in your reply is one EDIT to your notes, which are shown back to you every step. Every entry is numbered, like [3], and keeps that number for the whole trial. Write what you have worked out and what you mean to do: a way of working that paid off or wasted your time, what the river needed before apples appeared, what the other agents seem to be doing and whether to keep clear of them or follow them, what you will try next and why.
5.3 The four edits. add: a new entry in "text". replace: entry [id] becomes "text" -- use it the moment an entry turns out wrong or you can say it better, and that includes the lessons and plans you carried in from the last trial, which are the entries most likely to be wrong. delete: entry [id] is removed, for an entry that is wrong with nothing to put in its place. none: leave the notes as they are, which is right on most steps. A wrong entry left standing will mislead you on every step until you change it.
5.4 When a trial ends you are asked to rewrite your notes into a short list. THAT LIST IS ALL YOU KEEP of them: the step-by-step lines are dropped, and only what you put in the list is carried into the next trial. RULES YOU KNOW carries over on its own, so spend the list on how to play.
\end{lstlisting}
\begin{lstlisting}[style=prompt, caption={System prompt, OpenHarvest.}, label={lst:system_harvest}]
You are agent A, one of 5 agents (A, B, C, D, E) in a grid world.

1. YOUR OBJECTIVE
1.1 Score = the points the apples you eat pay. Only the points YOU earn count for you.
1.2 The trial starts with apples on every tree, and apples also grow back while it runs; whether an emptied cell grows one back depends on what is still hanging near it (2.3).
1.2b Apples come in several colours: each patch's colour is drawn at random, so two patches can have the same colour. What one pays depends on its colour and its ripeness stage. That is hidden and the same for every agent; you pick it up as you eat.
1.3 A trial lasts 400 steps. Then the orchard is re-laid with unripe apples on every tree and you go back to a spawn point. The colours, the places of the patches and the hidden rules (2.2, 2.3) stay the same for all 5 trials of the episode.
1.4 You are told only what YOU ate and what YOU see. You never see other agents' scores.

2. THE WORLD
2.1 An apple has a visible RIPENESS (unripe -> ripe -> over-ripe) that advances over time; an over-ripe apple eventually rots and vanishes, which counts as if nobody ate it.
2.2 You eat an apple by walking onto its cell, which empties that cell. The points it pays depend on its colour and its stage (hidden; you read them off the points you got), and the stage you eat it at also sets how fast that cell comes back (2.3).
2.3 REGROWTH. An empty tree cell can grow a new apple, which always starts unripe. This is the only way a trial gives you more apples than the ones it started with.
- A cell whose apple was EATEN grows a new one ONLY while apples are still hanging close to that cell. With no apples left around it, it NEVER grows one. A patch eaten bare therefore produces nothing for the rest of the trial, no matter how long you wait beside it.
- The more apples are still hanging CLOSE TO an empty cell, the sooner it grows a new one, and the drop is steep, not gradual: with several apples around it a cell usually comes back within the trial; with only two it takes many times longer; with a single one it rarely comes back at all; with none it never does. Only the apples near that cell count for it: apples hanging in another patch do nothing for it.
- How fast an eaten cell grows one also depends on its colour and on the ripeness stage its apple had when it was eaten. This part is hidden: watch the cells you empty.
- A cell whose apple ROTTED grows a new one by itself shortly afterwards, whatever is around it.
2.4 Moves are absolute compass moves and do not change your facing; only turning does. Ground and empty tree cells are walkable; walls and other agents block you. Walking onto an apple eats it.
2.5 ZAP fires a short beam ahead (the cell you face, the one beyond, and the two beside the faced cell). An agent hit by it is sent back to a spawn point and cannot act for the next 12 steps, so it loses those steps and the walk back as well. After you fire, your own beam needs a moment before it can fire again. Nobody gains or loses points from a zap, and it does nothing to apples.

3. HOW TO READ YOUR TURN
3.1 Positions are (row, col): row 0 is the NORTH edge and grows southward; col 0 is the WEST edge and grows eastward. North = row-1, south = row+1, east = col+1, west = col-1.
3.2 You see an 11x11 window that turns with you: 9 cells ahead of you, 1 cell behind and 5 to each side, so turning changes what you see. The map shows exactly those cells, always drawn north-up with columns labelled on top and rows on the left (ME sits one cell in from the side behind you); cells beyond the edge of the world are ##. ME = you, @B = agent B (every other agent is shown by its letter), ## = wall, .. = ground. An apple is its colour letter plus its stage letter: colour letters R = red, Y = yellow, G = green, C = cyan, P = purple; stage letters g = unripe, r = ripe, o = over-ripe (Yg = yellow unripe, Yr = yellow ripe, Yo = yellow over-ripe). An EMPTY tree cell is its colour letter in lower case plus an underscore, e.g. y_ = a yellow tree cell with no apple.
3.3 RULES YOU KNOW: what you have learned yourself: the points each colour and stage paid you; for cells you saw being emptied, how many steps they took to regrow (by colour and by the stage the apple was eaten at); how long apples took to ripen and to rot while you watched them; the rotting you have seen. YOUR NOTES: yours, and the only thing here that is -- what you have worked out about how to PLAY, and what you mean to do next. Nothing in it is checked or filled in for you. EARLIER TRIALS: one entry per finished trial: what you ate per colour and stage, and what you saw of the others. OTHER AGENTS: only what you have seen in your own view: where each agent was last seen and what it ate there. YOUR PLAN: your current goal and the goals you set before, kept for you. LAST STEP: what your last action did, the events in your view, and the earlier events of this trial. RECENT: your last steps. NOW also lists the patches you have seen, what the ones out of view looked like when you last saw them this trial, how long the apples you watched have been at their stage, and the emptied cells you are waiting on. ACTIONS NOW: what you can do right now and what each move lands on.
3.4 You know only what you see: other agents and patches exist beyond your view, but you get no information about them.

4. HOW TO ANSWER
4.1 Reply with ONE JSON object and nothing else:
{"why_this_move": "<up to 500 characters>",
 "mode": "harvest" | "wait" | "explore" | "travel" | "zap",
 "goal": {"type": "move_to" | "eat" | "wait" | "zap", "cell": [row, col], "why": "<up to 80 characters>"},
 "note": {"op": "none" | "add" | "replace" | "delete", "id": <entry number to replace or delete, else 0>, "text": "<the entry to add, or the new wording for [id]; \"\" otherwise>"},
 "action": <one integer from ACTIONS NOW>}
4.2 mode = what you are working on: harvest (going to eat an apple), wait (staying near apples for them to ripen or regrow), explore (testing something you do not know yet: a colour or stage you have not eaten, or a cell you emptied that you want to watch), travel (moving to another patch), zap (firing the beam at an agent).
4.3 goal = the sub-goal your action serves. Keep the SAME goal from step to step until it is done or an event makes it pointless; if you change it, say why in "why". For eat, cell = the apple's cell; for wait, cell = where you wait; for zap, cell = the cell of the agent you fire at.
4.4 Example, keeping the goal:
{"why_this_move": "I have not eaten a yellow ripe apple yet, so I do not know what it pays me. Yr at (1,4); I am at (3,4), two steps north.", "mode": "explore", "goal": {"type": "eat", "cell": [1, 4], "why": "find out what yellow ripe pays"}, "note": {"op": "none", "id": 0, "text": ""}, "action": 0}
4.5 Example, changing the goal after an event:
{"why_this_move": "The Cr at (6,2) I was walking to just got eaten by agent B. The next cyan ripe is at (6,4).", "mode": "harvest", "goal": {"type": "eat", "cell": [6, 4], "why": "previous target eaten by B"}, "note": {"op": "add", "id": 0, "text": "agent B also eats from the cyan patch"}, "action": 2}

5. YOUR NOTES
5.1 What you ate and what it paid, and how long the cells you watched took to grow back, are kept for you under RULES YOU KNOW, for the whole episode. You never have to write those down, and you should not: re-recording them only clutters your notes.
5.2 What is NOT kept for you is how to play. "note" in your reply is one EDIT to your notes, which are shown back to you every step. Every entry is numbered, like [3], and keeps that number for the whole trial. Write what you have worked out and what you mean to do: a way of eating that paid off or cost you later, how the patches behaved, what the other agents seem to be doing and whether to keep clear of them or follow them, what you will try next and why.
5.3 The four edits. add: a new entry in "text". replace: entry [id] becomes "text" -- use it the moment an entry turns out wrong or you can say it better, and that includes the lessons and plans you carried in from the last trial, which are the entries most likely to be wrong. delete: entry [id] is removed, for an entry that is wrong with nothing to put in its place. none: leave the notes as they are, which is right on most steps. A wrong entry left standing will mislead you on every step until you change it.
5.4 When a trial ends you are asked to rewrite your notes into a short list. THAT LIST IS ALL YOU KEEP of them: the step-by-step lines are dropped, and only what you put in the list is carried into the next trial. RULES YOU KNOW carries over on its own, so spend the list on how to play.
\end{lstlisting}
\begin{lstlisting}[style=prompt, caption={Example step prompt, OpenCleanup (gpt-6-sol, trial 3, step 120, agent B). Long lines are wrapped.}, label={lst:step_cleanup}]
Trial 3 of 5, step 120 of 400. This trial: you cleared 0 waste cells, you ate 6 apples.

RULES YOU KNOW
Orders you have fired with so far (what you watched them do):
  [0, 3] -> turned G2 (green-dark) into G1 (green-mid); no effect on G0 (green-light), G1 (green-mid)
  [1, 0] -> no effect on G0 (green-light), G1 (green-mid), G2 (green-dark)
  [1, 2] -> no effect on G0 (green-light), G1 (green-mid) (3x)
  [2, 1] -> no effect on G1 (green-mid), G2 (green-dark)
  [2, 3] -> no effect on G0 (green-light) (5x), G1 (green-mid)
  [3, 0] -> no effect on G0 (green-light) (2x)
  [3, 2] -> no effect on G1 (green-mid) (2x)
  [3, 3] -> no effect on G0 (green-light) (3x), G1 (green-mid)

YOUR NOTES (what you worked out about playing)
What you worked out by the end of trial 2:
  [1] Harvesting while others cleaned earned 10 apples in trial 1 and 15 in trial 2; my tool testing cleared no waste.
  [2] Apples appeared by step 65 after others had cleared about eight river cells; go to the orchard early.
  [3] C lightened green-dark waste, E lightened green-mid, and D cleared green-light in trial 2. Let them clean while I harvest.
  [4] From (2,3), facing south reveals orchard rows 1-11; wait there when no apple is visible rather than pacing.
  [5] Other agents clustered around rows 4-7; the southern orchard near (10,2) may offer less-contested apples.
  [6] The orchard extends to row 11; row 12 and columns 0-1 are blocked near its southern edge.

What you meant to try this trial:
  [7] Head to the orchard early and move promptly toward visible apples.
  [8] Watch from (2,3) when apples are scarce, then check the southern orchard if nearby apples are contested.
  [9] If apples stop spawning, check river progress before spending time testing tools.

Notes you have added since:
  [10] t18: Try (10,3) facing north as a southern lookout: it should show orchard rows 1-11 and give a head start on southern apples.
  [11] t48: t68: Apples appeared after I made [0,3]. Prioritize visible apples over cleaning; keep [0,3] in hand in case spawning stalls again. (revised t69)
  [12] t116: At step 116 the orchard had no visible apples and all four other agents were in it; check river progress when spawning stalls. Keep [0,3] for G2 waste.

EARLIER TRIALS
Trial 1: you cleared 0 waste cells, ate 10 apples, fired 8 times, spent 15 steps in or next to the river; first working tool in hand at step 48.
Trial 1, other agents: A seen 232 steps, you saw it fire 10 time(s): cleared 7 G0 (green-light); turned 4 G1 (green-mid) into G0 (green-light); 3 shot(s) hit waste with no effect; C seen 252 steps, you saw it fire 7 time(s): turned 2 G2 (green-dark) into G1 (green-mid); 5 shot(s) hit waste with no effect; D seen 190 steps, you saw it fire 13 time(s): cleared 4 G0 (green-light); 6 shot(s) hit waste with no effect; 3 shot(s) with no waste under the beam in your view; E seen 228 steps, you saw it fire 13 time(s): turned 1 G1 (green-mid) into G0 (green-light); 10 shot(s) hit waste with no effect; 2 shot(s) with no waste under the beam in your view; 11 river cells in your view were cleared by others.
Trial 2: you cleared 0 waste cells, ate 15 apples, fired 2 times, spent 8 steps in or next to the river; first working tool in hand at step 63.
Trial 2, other agents: A seen 273 steps, you saw it fire 5 time(s): 1 shot(s) hit waste with no effect; 4 shot(s) with no waste under the beam in your view; C seen 250 steps, you saw it fire 6 time(s): turned 4 G2 (green-dark) into G1 (green-mid); 3 shot(s) with no waste under the beam in your view; D seen 305 steps, you saw it fire 12 time(s): cleared 12 G0 (green-light); 1 shot(s) hit waste with no effect; 4 shot(s) with no waste under the beam in your view; E seen 318 steps, you saw it fire 12 time(s): turned 11 G1 (green-mid) into G0 (green-light); 2 shot(s) hit waste with no effect; 2 shot(s) with no waste under the beam in your view; 11 river cells in your view were cleared by others.

OTHER AGENTS (what you have seen yourself)
Agent A: last seen now at (11, 5) facing north; seen 50 times this trial.
Agent C: last seen 1 steps ago at (9, 1) facing east; seen 58 times this trial.
Agent D: last seen now at (2, 3) facing north; seen 74 times this trial; you saw it fire 2 time(s): cleared 4 G0 (green-light).
Agent E: last seen now at (9, 2) facing east; seen 93 times this trial.
River cells cleared in your view but not by you this trial: 4.

YOUR PLAN
Current goal #76 (set step 116, 4 steps ago): move_to (8, 10) -- "reach the river bank to check for G2 waste". Distance now 6 (was 9).
Earlier goals this trial: #71 wait (10, 3) -- done at step 111 | #72 wait (10, 3) -- done at step 112 | #73 wait (10, 3) -- done at step 113 | #74 wait (10, 3) -- done at step 114 | #75 wait (10, 3) -- done at step 115

LAST STEP
Last step (119): you chose 2 (east) -> moved (10, 6) -> (10, 7)
Events: Agent C left your view.

RECENT STEPS
t105 east -> (7, 3)
t106 south -> (8, 3)
t107 south -> (9, 3)
t108 south -> (10, 3)
t109 south -> (11, 3)
t110 north -> (10, 3)
t111 turn left -> facing east
t112 turn left -> facing north
t113 stay
t114 stay
t115 stay
t116 east -> (10, 4)
t117 east -> (10, 5)
t118 east -> (10, 6)
t119 east -> (10, 7)

NOW
You are agent B, at (10, 7), facing north.
You hold what your pick-ups [0, 3] gave you. Your window (last two pick-ups, oldest first): [0, 3]; your next pick-up x makes it [3, x].
In view: dominant hue green (light 1, mid 3, dark 3); river cells you can see: 7 waste, 8 water; apples 0; other agents: A at (11, 5), D at (2, 3), E at (9, 2).
Stations in view: T0 at (4, 6), (10, 8); T1 at (6, 6), (8, 8); T2 at (6, 8), (8, 6); T3 at (4, 8), (10, 6).
Stations you have seen (same places every trial): T0 at (4, 6), (10, 8); T1 at (6, 6), (8, 8); T2 at (6, 8), (8, 6); T3 at (4, 8), (10, 6).
River seen (same place every trial): rows 0-8, cols 11-15 (37 cells seen); orchard seen: rows 0-11, cols 0-5 (48 cells seen).
Map:
     2  3  4  5  6  7  8  9 10 11 12
r 1 .. .. .. .. .. .. .. .. .. .. ~~
r 2 ,, @D .. .. .. .. .. .. .. G1 ~~
r 3 ,, .. .. .. .. .. .. .. .. ~~ G2
r 4 ,, ,, .. .. T0 .. T3 .. .. G2 ~~
r 5 ,, ,, ,, .. .. .. .. .. .. ~~ ~~
r 6 ,, ,, ,, .. T1 .. T2 .. .. ~~ G1
r 7 ,, ,, ,, .. .. .. .. .. .. G1 G2
r 8 ,, .. ,, .. T2 .. T1 .. .. ~~ G0
r 9 @E ,, ,, .. .. .. .. .. .. .. ..
r10 ,, ,, .. .. T3 ME T0 .. .. .. ..
r11 ,, ,, ,, @A .. .. .. .. .. .. ..

ACTIONS NOW
0 north | 1 south | 2 east | 3 west | 4 turn left | 5 turn right | 6 fire (nothing in the beam) | 7 stay

Reply with the JSON only.
\end{lstlisting}
\begin{lstlisting}[style=prompt, caption={Example step prompt, OpenHarvest (gpt-6-sol, trial 1, step 40, agent B). Long lines are wrapped.}, label={lst:step_harvest}]
Trial 1 of 5, step 40 of 400. This trial: you ate 5 apples for 3.2 points.

RULES YOU KNOW
What you ate and what it paid you: red: unripe -> 1 points (x1); cyan: unripe -> 1 points (x1), ripe -> 0.25 points (x1), over-ripe -> 0.5 points (x2).
Regrowth you watched (cells you saw emptied; steps until a new apple appeared, <= when you looked away; a cell only grows one while apples hang around it): red cells eaten at unripe: 3 not regrown yet this trial (still empty up to 8 steps after being emptied; apples around: 4-5 when emptied, 3-5 at your last look)
green cells eaten at over-ripe: 1 not regrown yet this trial (still empty up to 1 steps after being emptied; apples around: 4 when emptied, 3 at your last look)
cyan cells eaten at unripe: 1 not regrown yet this trial (still empty up to 12 steps after being emptied; apples around: 2 when emptied, 1 at your last look)
cyan cells eaten at ripe: 2 not regrown yet this trial (still empty up to 13 steps after being emptied; apples around: 3-4 when emptied, 2 at your last look)
cyan cells eaten at over-ripe: 2 not regrown yet this trial (still empty up to 2 steps after being emptied; apples around: 1-2 when emptied, 1-2 at your last look)
Ripening you watched (apples you kept in view from the moment they appeared): ripe -> over-ripe after 16, 17, 18 steps.
Apples you saw rot: none.

YOUR NOTES (what you worked out about playing)
Notes you have added since:
  [1] t25: Cyan ripe pays less than unripe; avoid clearing the patch while testing over-ripe fruit, since hanging apples support regrowth.

OTHER AGENTS (what you have seen yourself)
Agent A: last seen 29 steps ago at (5, 11) facing south; seen 11 times this trial; ate 1 in your view (red unripe 1); fired 1 times in your view.
Agent C: last seen 7 steps ago at (14, 16) facing east; seen 14 times this trial; ate 1 in your view (red unripe 1).
Agent D: last seen now at (12, 16) facing north; seen 19 times this trial; ate 1 in your view (cyan ripe 1); fired 2 times in your view; sent you back to a spawn point 2 times.
Agent E: last seen now at (14, 20) facing east; seen 5 times this trial; ate 2 in your view (green over-ripe 2); sent you back to a spawn point 1 times.

YOUR PLAN
Current goal #9 (set step 36, 4 steps ago): eat (14, 20) -- "test green over-ripe payoff". Distance now 5 (was 8).
Earlier goals this trial: #1 eat (4, 10) -- done at step 9 | #2 zap (4, 11) -- done at step 10 | #3 eat (6, 19) -- abandoned at step 18 ("test the payoff of ripe cyan") | #4 eat (6, 20) -- done at step 23 | #5 eat (6, 19) -- done at step 24 | #6 eat (6, 21) -- abandoned at step 30 ("wait for adjacent cyan apple to become over-ripe") | #7 wait (5, 21) -- abandoned at step 32 ("test cyan over-ripe payoff") | #8 eat (7, 21) -- done at step 35

LAST STEP
Last step (39): you chose 1 (south) -> moved (9, 20) -> (10, 20).
Events: Agent E ate a green over-ripe apple at (14, 20).
Earlier events this trial (last 14): t7 Agent C ate a red unripe apple at (6, 12).; t9 Agent D came into your view at (8, 19).; t10 Regrowth in your view: 1 cyan (you did not see that cell being emptied).; t11 Agent A ate a red unripe apple at (5, 11).; t11 Agent D left your view.; t12 Agent A, C left your view.; t15 Agent C came into your view at (9, 13).; t17 Agent C left your view.; t24 Agent D came into your view at (9, 19).; t28 Agent D ate a cyan ripe apple at (7, 20).; t33 Agent C came into your view at (14, 16).; t34 Agent C left your view.; t36 Agent E came into your view at (14, 16).; t39 Agent E ate a green over-ripe apple at (14, 19).

RECENT STEPS
t25 E -> (6, 20)
t26 stay
t27 N -> (5, 20)
t28 E -> (5, 21)
t29 stay
t30 stay
t31 turnR
t32 W -> (5, 20)
t33 S -> (6, 20)
t34 E -> (6, 21) ate Co +0.5
t35 S -> (7, 21) ate Co +0.5
t36 W -> (7, 20)
t37 S -> (8, 20)
t38 S -> (9, 20)
t39 S -> (10, 20)

NOW
You are agent B, at (10, 20), facing south.
Patches in view: green rows 11-13 cols 15-15: 0 apples (unripe 0, ripe 0, over-ripe 0), 3 empty tree cells; green rows 13-15 cols 19-21: 4 apples (unripe 0, ripe 1, over-ripe 3), 1 empty tree cells.
Other agents in view: D at (12, 16), E at (14, 20).
Patches you have seen (same places every trial): red rows 4-8 cols 8-12; cyan rows 6-8 cols 19-21; green rows 11-13 cols 15-16; green rows 13-15 cols 19-21.
Tree cells out of view, as you last saw them this trial: red rows 4-8 cols 8-12 (13 tree cells seen so far): at step 15 you saw 1 of them: 0 apples (unripe 0, ripe 0, over-ripe 0) and 1 empty cell; the other 12 you last saw at steps 8-14: 10 apples (unripe 10, ripe 0, over-ripe 0) and 2 empty cells; cyan rows 6-8 cols 19-21 (6 tree cells seen so far): at step 39 you saw 1 of them: 1 apple (unripe 0, ripe 0, over-ripe 1) and 0 empty cells; the other 5 you last saw at steps 37-38: 0 apples (unripe 0, ripe 0, over-ripe 0) and 5 empty cells; green rows 11-13 cols 15-16 (4 tree cells seen so far, 3 in view now): at step 39 you saw the rest: 0 apples (unripe 0, ripe 0, over-ripe 0) and 1 empty cell; green rows 13-15 cols 19-21 (6 tree cells seen so far, 5 in view now): at step 39 you saw the rest: 1 apple (unripe 0, ripe 0, over-ripe 1) and 0 empty cells.
Apples in view whose clock you know: (13, 19) green ripe since step 35.
Cells you are waiting on (you saw them emptied this trial and have not seen them regrow): (14, 20) green eaten over-ripe at step 40, empty for 0 steps so far; (14, 19) green eaten over-ripe at step 39, empty for 1 steps so far; (7, 21) cyan eaten over-ripe at step 36, still empty when you last saw it at step 38; (6, 21) cyan eaten over-ripe at step 35, still empty when you last saw it at step 37; (7, 20) cyan eaten ripe at step 28, still empty when you last saw it at step 38; (6, 19) cyan eaten unripe at step 25, still empty when you last saw it at step 37; (6, 20) cyan eaten ripe at step 24, still empty when you last saw it at step 37; (5, 11) red eaten unripe at step 11, still empty when you last saw it at step 14.
Map:
    15 16 17 18 19 20 21            
r 9 .. .. .. .. .. .. .. ## ## ## ##
r10 .. .. .. .. .. ME .. ## ## ## ##
r11 g_ .. .. .. .. .. .. ## ## ## ##
r12 g_ @D .. .. .. .. .. ## ## ## ##
r13 g_ .. .. .. Gr .. .. ## ## ## ##
r14 .. .. .. .. g_ @E .. ## ## ## ##
r15 .. .. .. .. Go Go Go ## ## ## ##
r16 ## ## ## ## ## ## ## ## ## ## ##
r17 ## ## ## ## ## ## ## ## ## ## ##
r18 ## ## ## ## ## ## ## ## ## ## ##
r19 ## ## ## ## ## ## ## ## ## ## ##

ACTIONS NOW
0 north -> (9, 20): ground | 1 south -> (11, 20): ground | 2 east -> (10, 21): ground | 3 west -> (10, 19): ground | 4 turn left | 5 turn right | 6 zap (no agent in the beam) | 7 stay

Reply with the JSON only.
\end{lstlisting}

\FloatBarrier
% ---------------------------------------------------------------------------------------------------
\section{Models and Inference}
\label{app:models}

\Cref{tab:app_models} lists the models and their inference settings, which are the same in both environments. Claude Opus 5.5 declined structured output in our setting, so its replies are parsed from free text. DeepSeek V4.1 Flash runs without thinking. We observed that enabling the thinking mode for DeepSeek V4.1 Flash frequently leads to overthinking and excessively long reasoning traces. Therefore, we disable thinking mode in our experiments.
% TODO verify the sampling temperature used for Gemma 4 31B and Qwen3.8-27B (no explicit value in the run config).

\begin{table}[htbp]
\centering
\small
\caption{Models and inference settings (identical in both environments). Qwen3.5-4B samples with top-$p$ 0.95, top-$k$ 20 and presence penalty 1.5, and its thinking is capped at 3000 tokens.}
\label{tab:app_models}
\begin{tabular}{llllrl}
\toprule
Model & Access & Thinking & Temp.  & Output \\
\midrule
Claude Opus 5.5 & API & medium & default  & parsed text \\
gpt-6-sol & API & built-in & default  & JSON schema \\
gpt-6-luna & API & built-in & default  & JSON schema \\
Gemma 4 31B & vLLM & on & default  & parsed text \\
GLM-5.3-Flash & API & low & 0.2  & JSON mode \\
Qwen3.8-27B & vLLM & medium & default  & parsed text \\
DeepSeek V4.1 Flash & API & off & 0.2  & JSON mode \\
Qwen3.5-4B & vLLM & on & 1.0  & parsed text \\
\bottomrule
\end{tabular}
\end{table}

Table~\ref{tab:experiments} summarizes the configuration of each experiment, including the evaluated models, environments, number of seeds, and number of trials.

\begin{table}[h]
\centering
\small
\caption{All experiments use teams of five agents, and all trials last at most 400 steps; an OpenHarvest trial ends early once no apple remains. \emph{Env.}: OpenCleanup (C) and/or OpenHarvest (H). \emph{Seeds}: seeds per environment; each seed samples one latent configuration (map layout and hidden rules), and the same seeds are shared across experiments. For the Schelling diagram, seeds are counted per team composition.}
\label{tab:experiments}
\begin{tabular}{lllcc}
\toprule
Experiment & Models & Env. & Seeds & Trials \\
\midrule
Homogeneous teams  & all 8       & C, H & 3    & 5 \\
Mixed-model arena  & 5, one each & C, H & 3    & 5 \\
Schelling diagram  & gpt-6-luna& C, H & 3 & 5 \\
Scripted greedy    & --          & H    & 3    & 5 \\
\bottomrule
\end{tabular}
\end{table}

% \Cref{tab:app_cost} reports the inference cost per run. An OpenCleanup run always takes 10{,}020 calls (5 agents $\times$ 5 trials $\times$ 400 steps, plus 20 note rewrites). OpenHarvest runs are shorter and vary in length because trials stop early once the orchard is dead (\cref{app:harness}).

% \input{appendix/tables/tab_cost}
% \input{appendix/tables/tab_reliability}

\FloatBarrier
% ------------------------------------------------------------------------------

% ------------------------------------------------------------------------------
% ---------------------------------------------------------------------------------------------------
\section{Additional Results}
\label{app:results}

\subsection{Schelling Diagrams}
\label{app:schelling}

\textbf{Setup.}
To verify that both environments induce sequential social dilemmas, we compute Schelling diagrams following \citet{leibo2017multi}. We instantiate cooperating and defecting policies with GPT-6-Luna. 
Each agent receives one of two roles, stated as item 1.5 (\textsc{Your role}) of its system prompt (Listing~\ref{lst:Schelling}). Apart from this item, the prompts of the two roles are identical: item 1.1 states only the scoring rule, and the environment pays every agent individually in both cases. The defector role asks only for self-interest and does not forbid cleaning. Agents are not told the other agents' roles.

\begin{lstlisting}[style=prompt, caption={System prompt for the Schelling diagram runs. Each agent sees only its own role.}, label={lst:Schelling}]
OpenCleanup, cooperator:
1.5 YOUR ROLE: You care about the whole group, not only about yourself. You are willing
    to give up some of your own apples, and to spend your time on work that helps
    everyone, if it lets the others eat more. The group's total matters more to you
    than your own score.

OpenCleanup, defector:
1.5 YOUR ROLE: You care only about your own score. Your goal is to eat as many apples
    as you can yourself. What the other agents eat does not matter to you.

OpenHarvest, cooperator:
1.5 YOUR ROLE: You care about the whole group, not only about yourself. You are willing
    to give up some of your own points, and to spend your time on work that helps
    everyone, if it lets the others earn more. The group's total matters more to you
    than your own score.

OpenHarvest, defector:
1.5 YOUR ROLE: You care only about your own score. Your goal is to earn as many points
    as you can yourself. What the other agents earn does not matter to you.
\end{lstlisting}

For $k = 0, \dots, 5$ cooperators, all five agents are played by GPT-6-Luna, and which agents are cooperators rotates across three seeds. Each seed is one episode of five 400-step trials, following the same multi-trial protocol as the main experiments; agents keep their roles for the whole episode. All other parts of the harness and the inference settings are identical to the homogeneous evaluation, and the hidden rules are not revealed. % TODO: 如果用了公开规则，改掉这半句
Returns are averaged over the five trials.

\begin{figure}[h]
\centering
\begin{subfigure}[b]{0.48\linewidth}
  \includegraphics[width=\linewidth]{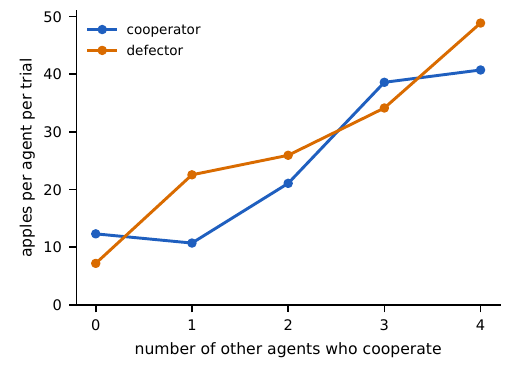}
  \caption{OpenCleanup}
\end{subfigure}\hfill
\begin{subfigure}[b]{0.48\linewidth}
  \includegraphics[width=\linewidth]{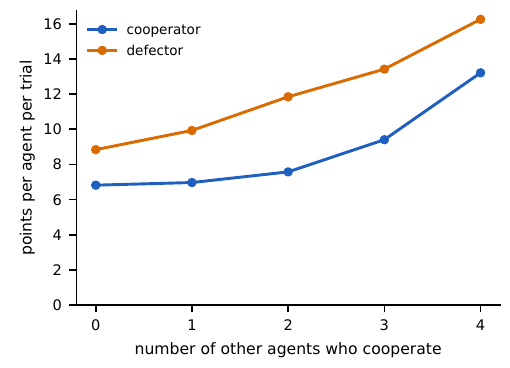}
  \caption{OpenHarvest}
\end{subfigure}
\caption{Schelling diagrams, averaged over five trials. Blue: return of a cooperator, $C(x)$; orange: return of a defector, $D(x)$; $x$ is the number of other agents that are cooperative.}
\label{fig:schelling}
\end{figure}

\textbf{Reading the diagram.}
Let $C(x)$ and $D(x)$ denote the mean return of a cooperator and a defector when $x$ of its four co-players are cooperators; in the notation of Section~\ref{sec:difine}, $C(x) = R_c(x+1)$ and $D(x) = R_d(x)$. Comparing the two curves at the same $x$ shows whether an agent earns more by cooperating or by defecting, given the behavior of its co-players. Both environments satisfy the conditions of a sequential social dilemma (Figure~\ref{fig:schelling}). In OpenHarvest, defecting pays more for every number of cooperating co-players, so both fear and greed hold. In OpenCleanup, greed holds, since exploiting four cooperators pays more than mutual cooperation, but fear does not: a lone cooperator among defectors earns more than a defector under mutual defection, since apples grow only if someone cleans. 

\subsection{Total Cleaning in OpenCleanup}
\label{app:total-cleaning}

\begin{figure}[h]
\centering
\begin{subfigure}[b]{0.4\linewidth}
  \includegraphics[width=\linewidth]{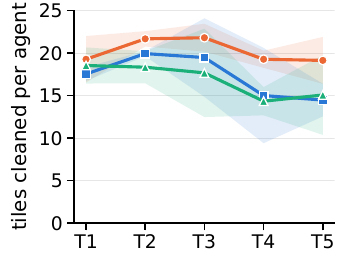}
  \caption{Closed-weight models}
\end{subfigure}\hfill
\begin{subfigure}[b]{0.4\linewidth}
  \includegraphics[width=\linewidth]{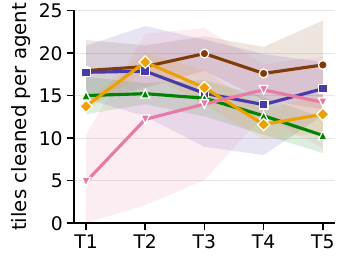}
  \caption{Open-weight models}
\end{subfigure}

\vspace{2pt}
\includegraphics[width=0.9\linewidth]{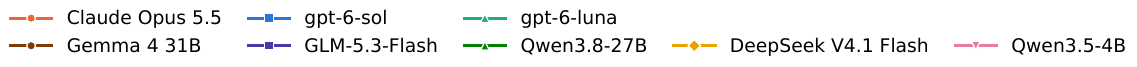}
\caption{Waste tiles cleared per agent (team total divided by five) in each OpenCleanup trial. Shaded bands: $\pm 1$ s.d.\ across 3 seeds.}
\label{fig:total-cleaning}
\end{figure}

A falling effective number of cleaners can reflect either free-riding or a division of labor. To tell them apart, Figure~\ref{fig:total-cleaning} shows the team's total cleaning, counted as in \Cref{fig:dilemma_cleaners_closed,fig:dilemma_cleaners_open}: waste tiles cleared, summed over the five agents. We compare trials 2 and 5, since in trial 1 agents are still learning the tool recipes.

Claude Opus 5.5 and Gemma 4 31B keep cleaning high across trials, with only a slight decline for Claude Opus 5.5, and their returns hold or rise, consistent with specialization. GPT-6-Sol, GPT-6-Luna, Qwen3.8-27B and DeepSeek-V4.1-Flash clean noticeably less in later trials and earn less, consistent with free-riding. GLM-5.3-Flash shows a slight decline similar to Claude Opus 5.5, but its returns fall; Qwen3.5-4B cleans more over time as it learns the recipes.

\FloatBarrier
\subsection{Why OpenCleanup Returns Jump After the First Trial}
\label{app:mechanism}

\Cref{tab:app_mechanism} shows what changes between the first trial and the later ones. In trial 1, agents spend a quarter to a third of their steps testing pick-up orders, and most teams clear their first cell only after 60 to 70 steps. From trial 2 on, the recipes are carried over in RULES YOU KNOW and the layout in the notes. Exploration almost stops, the first cell is cleared within the first 10 to 25 steps for most models, cleaning in the first 50 steps rises several-fold, and apples appear much earlier. Claude Opus 5.5 stands out already in trial 1, when it clears its first cell after about 20 steps. From trial 2 on, its teams do nearly twice as much early cleaning as the next model, and their first apples appear within about 50 steps.

\begin{table}[htbp]
\centering
\small
\caption{OpenCleanup: what changes after the first trial. Share of steps in \emph{explore} mode, the step of the team's first cleared cell and first eaten apple, and cells cleared or lightened in the first 50 steps; trial 1 vs.\ the mean of trials 2--5.}
\label{tab:app_mechanism}
\begin{tabular}{lrrrrrrrr}
\toprule
 & \multicolumn{2}{c}{Explore (\%)} & \multicolumn{2}{c}{First clear} & \multicolumn{2}{c}{First apple} & \multicolumn{2}{c}{Cells, $t<50$} \\
\cmidrule(lr){2-3}\cmidrule(lr){4-5}\cmidrule(lr){6-7}\cmidrule(lr){8-9}
Model & T1 & T2--5 & T1 & T2--5 & T1 & T2--5 & T1 & T2--5 \\
\midrule
Claude Opus 5.5 & 31 & 1 & 21 & 16 & 157 & 47 & 7 & 37 \\
gpt-6-sol & 30 & 3 & 58 & 11 & 145 & 74 & 6 & 18 \\
gpt-6-luna & 33 & 4 & 67 & 14 & 142 & 94 & 3 & 16 \\
Gemma 4 31B & 27 & 3 & 59 & 17 & 170 & 75 & 6 & 21 \\
GLM-5.3-Flash & 27 & 2 & 70 & 17 & 237 & 153 & 1 & 21 \\
Qwen3.8-27B & 22 & 1 & 68 & 24 & 209 & 179 & 4 & 15 \\
DeepSeek V4.1 Flash & 33 & 11 & 60 & 40 & 299 & 188 & 6 & 10 \\
Qwen3.5-4B & 15 & 10 & 324 & 63 & 390 & 257 & 0 & 8 \\
\bottomrule
\end{tabular}
\end{table}

\subsection{Tool Use}
\label{app:tools}

Agents that know more recipes pick up tools more often. Within the same run and trial, an agent's number of known recipes correlates with its pick-ups (median Spearman $\rho = 0.71$ over 108 run--trials) and tool switches ($\rho = 0.82$). This does not make them less efficient. In trials 2--5, agents that know two or three recipes clean about 15.6 cells per 100 steps spent crafting or cleaning, against 9.3 for agents that know one. Knowing several recipes lets an agent finish a cell (dark $\to$ mid $\to$ light $\to$ water) by itself, which requires switching tools. \Cref{tab:app_tools} shows that most models need close to the minimum of two pick-ups per working tool. Claude Opus 5.5 needs fewer than two because it exploits the sliding window: one extra pick-up can turn one recipe into the next. DeepSeek V4.1 Flash and Qwen3.5-4B often drop a working tool by an extra pick-up and craft it again. They spend the most crafting steps per cleaned cell: DeepSeek up to about twice as many as the other models, and Qwen3.5-4B about five times as many.

\begin{table}[htbp]
\centering
\small
\caption{OpenCleanup tool use in trials 2--5: pick-ups per agent per trial, pick-ups per working tool obtained (2 is the minimum unless a sliding window completes a recipe with one pick-up), share of working tools that re-crafted the tool the agent had just lost, and steps spent crafting per cell cleared or lightened.}
\label{tab:app_tools}
\begin{tabular}{lrrrr}
\toprule
Model & Pick-ups & Pick-ups / tool & Re-crafts (\%) & Craft steps / cell \\
\midrule
Claude Opus 5.5 & 6.4 & 1.9 & 2 & 1.58 \\
gpt-6-sol & 5.6 & 2.2 & 5 & 1.36 \\
gpt-6-luna & 5.1 & 2.6 & 19 & 2.09 \\
Gemma 4 31B & 5.3 & 2.3 & 4 & 1.71 \\
GLM-5.3-Flash & 6.3 & 2.5 & 9 & 2.00 \\
Qwen3.8-27B & 4.4 & 2.2 & 4 & 2.08 \\
DeepSeek V4.1 Flash & 15.8 & 4.1 & 31 & 3.35 \\
Qwen3.5-4B & 40.1 & 20.8 & 53 & 10.15 \\
\bottomrule
\end{tabular}
\end{table}

\subsection{Knowing the Payoffs Versus Using Them}
\label{app:use}

\begin{figure}[htbp]
    \centering
    \includegraphics[width=0.45\linewidth]{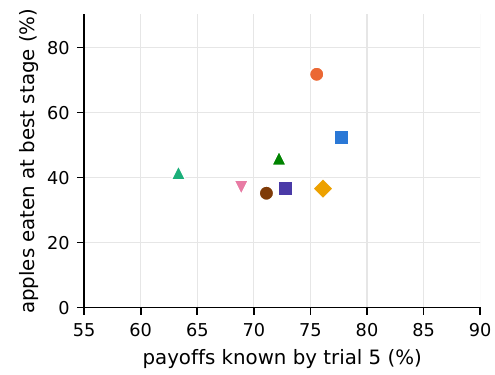}\\[2pt]
    \includegraphics[width=0.9\linewidth]{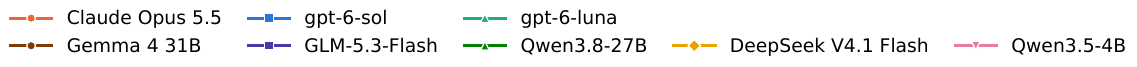}
    \caption{OpenHarvest self-play: share of the payoffs known by the end of trial 5 against the share of apples eaten at their best-paying stage.}
    \label{fig:app_knowing_vs_using}
\end{figure}

By trial 5, most models know a similar share of the OpenHarvest payoffs, yet their returns differ widely (Table~\ref{tab:app_use}). The difference is in how they eat. Claude Opus 5.5 takes almost three quarters of its apples at the stage that pays one point for their color, while the other models do so for a third to a half, and eat 38--55\% of their apples before that stage (\Cref{fig:app_knowing_vs_using}). Claude therefore earns more points per apple, in addition to keeping its orchard alive longer.

\begin{table}[htbp]
\centering
\small
\caption{OpenHarvest self-play: knowing the payoffs versus using them. Payoffs known by the end of trial 5; share of eaten apples taken at their best-paying stage and before it; mean points per apple; mean per-agent return.}
\label{tab:app_use}
\begin{tabular}{lrrrrr}
\toprule
Model & Known (\%) & Best stage (\%) & Premature (\%) & Points / apple & Return \\
\midrule
Claude Opus 5.5 & 76 & 72 & 20 & 0.84 & 13.5 \\
gpt-6-sol & 78 & 52 & 38 & 0.73 & 10.2 \\
gpt-6-luna & 63 & 42 & 49 & 0.66 & 8.9 \\
Gemma 4 31B & 71 & 35 & 55 & 0.61 & 8.3 \\
GLM-5.3-Flash & 73 & 37 & 53 & 0.63 & 8.6 \\
Qwen3.8-27B & 72 & 46 & 41 & 0.69 & 9.3 \\
DeepSeek V4.1 Flash & 76 & 37 & 55 & 0.62 & 8.9 \\
Qwen3.5-4B & 69 & 37 & 51 & 0.61 & 8.4 \\
\bottomrule
\end{tabular}
\end{table}

\FloatBarrier
% ---------------------------------------------------------------------------------------------------
\section{Agents' Own Notes}
\label{app:notes}

\Cref{tab:app_notes} quotes lessons the agents wrote for themselves at the end of a trial (Appendix~\ref{app:harness}). Letters refer to agents; in self-play all five agents are copies of the same model.

\begin{table}[htbp]
\centering
\small
\caption{Lessons written by agents at the end of a trial, quoted verbatim. Source: model (or environment, in the arena) / seed / agent / trial after which the lesson was written.}
\label{tab:app_notes}
\begin{tabularx}{\linewidth}{>{\raggedright\arraybackslash}p{4.1cm}>{\raggedright\arraybackslash}X}
\toprule
Source & Lesson \\
\midrule
\multicolumn{2}{l}{\emph{OpenCleanup self-play: free-riding}} \\
gpt-6-luna / 908 / A / T2 & ``Prioritizing the orchard paid off: I ate 39 apples without firing or entering the river.'' \\
 & ``I can rely on other agents' cleanup while focusing on harvesting; 25 river cells were cleared in my view this trial.'' \\
gpt-6-luna / 908 / C / T2 & ``I cleared 49 cells and spent 114 steps in or beside the river, but ate no apples; the growth threshold is still unknown.'' \\
gpt-6-sol / 906 / A / T2 & ``Harvesting while others cleaned yielded 44 apples in trial 2, versus 12 in trial 1, without clearing waste myself.'' \\
gpt-6-sol / 906 / B / T2 & ``C, D and E cleared or lightened cyan waste; I cleared none despite eight shots. Let them handle most cleaning while I harvest.'' \\
GLM-5.3-Flash / 906 / A / T2 & ``Trial 3: pick up 2 then 0 fast, clear row-9/10 C2 belt, then be in orchard by mid-trial.'' (plan; the agent cleared no waste in trial 3) \\
\midrule
\multicolumn{2}{l}{\emph{OpenCleanup self-play: division of labor}} \\
Claude Opus 5.5 / 906 / A / T1 & ``Others (C, D) lighten C2$\to$C1$\to$C0 and B, D, E clear C0. My role is sweeping the C0 they leave behind.'' \\
 & ``Late in the trial, harvest the east orchard (cols 12-15). I got 20 apples from harvesting late, so harvesting pays more than extra cleaning.'' \\
\midrule
\multicolumn{2}{l}{\emph{OpenHarvest self-play: restraint}} \\
Claude Opus 5.5 / 907 / A / T1 & ``Cyan was eaten bare by D and me by t64 and was then dead for the whole trial. Never take the last 3 of a patch.'' \\
\midrule
\multicolumn{2}{l}{\emph{Arena (Claude Opus 5.5's agent among four other models)}} \\
OpenCleanup / 907 / E / T3 & ``Others lighten waste: B does G2$\to$G1, A and C do G1$\to$G0 (A also clears G0). D never cleans. My best role is clearing G0 with [0,1].'' \\
OpenHarvest / 906 / A / T2 & ``Rivals strip every patch by about t50-65, even the west cyan, so there is almost no regrowth. Apples held back mostly go to others.'' \\
OpenHarvest / 907 / E / T4 & ``Red gets crowded by B and D by t8-12 and is stripped by t18. Leaving apples there to regrow is futile. Eat fast.'' \\
OpenHarvest / 908 / D / T2 & ``E is the main stripper: south purple, then cyan at \textasciitilde t50, then the centre. Zapping E once or twice did not stop it.'' \\
\bottomrule
\end{tabularx}
\end{table}

\FloatBarrier

% \begin{figure}[t]
%     \centering

%     \begin{subfigure}{\linewidth}
%         \centering
%         \includegraphics[width=\linewidth]
%         {figures/llm_agents_pipeline/llm_agent_pipeline_paper_cleanup_2026-09-21.pdf}
%         \caption{Structure of the Cleanup Discovery environment.}
%         \label{fig:cleanup-structure}
%     \end{subfigure}

%     \vspace{0.5em}

%     \begin{subfigure}{\linewidth}
%         \centering
%         \includegraphics[width=\linewidth]
%         {figures/llm_agents_pipeline/llm_agent_harvest_pipeline_paper_2026-09-21.pdf}
%         \caption{Fruit dynamics in the Harvest Discovery environment.}
%         \label{fig:harvest-rules}
%     \end{subfigure}

%     \caption{Overview of the environments in OpenSocialJax.}
%     \label{fig:environment-overview}
% \end{figure}

\end{document}